\documentclass[10pt,journal,compsoc]{IEEEtran}

\usepackage{multirow}
\usepackage{amsthm,amsmath}
\usepackage{caption}
\usepackage{xcolor,colortbl}
\usepackage{subcaption}
\usepackage[ruled,linesnumbered]{algorithm2e}
\usepackage{enumitem}
\usepackage{tcolorbox}
\usepackage{balance}
\usepackage{url}
\usepackage{booktabs}
\usepackage{amssymb}
\usepackage{tipa} %
\newcommand{\ipa}{\textipa}

\usepackage{amsmath,amssymb,amsfonts}
\usepackage[ruled,linesnumbered]{algorithm2e}
\usepackage{algorithmic}
\algsetup{linenosize=\small}
\SetKwInput{KwInput}{Input}                
\SetKwInput{KwOutput}{Output}              

\newcommand\ans[1]{
\noindent 
\fcolorbox{green!40!black}{green!5}{\noindent 
 \parbox{0.98\columnwidth}{\noindent  #1}}\\
}

\definecolor{mGreen}{rgb}{0,0.6,0}
\definecolor{mGray}{rgb}{0.5,0.5,0.5}
\definecolor{mPurple}{rgb}{0.58,0,0.82}
\definecolor{backgroundColour}{rgb}{0.95,0.95,0.92}

\definecolor{flatgreen}{HTML}{b7f4d8}
\definecolor{flatred}{HTML}{ffcd02}
\def \setGoodColor {\cellcolor{flatgreen}}
\def \setBadColor {\cellcolor{flatred}}

\usepackage{enumitem}
\usepackage{amsmath,amssymb}

\ifCLASSOPTIONcompsoc
  \usepackage[nocompress]{cite}
\else
  \usepackage{cite}
\fi

\ifCLASSINFOpdf
\else
\fi

\begin{document}

\def \toolname{\textsc{Prophet }}
\def \toolnamenospace{\textsc{Prophet}}

\def \numbermodel{$3,024$ }

\newcommand{\blue}[1]{\textcolor{blue}{#1}}
\newcommand{\zy}[1]{\mynote{Zhou}{#1}}
%
\title{Prioritizing Speech Test Cases}
%
%
%
%

\author{Zhou~Yang, Jieke~Shi, Muhammad~Hilmi~Asyrofi, Bowen~Xu, Xin Zhou, DongGyun Han, and~David~Lo~\IEEEmembership{Fellow,~IEEE}
\IEEEcompsocitemizethanks{
\IEEEcompsocthanksitem Z. Yang, J. Shi, B. Xu, X. Zhou, D. Lo are
with the School of Computing and Information Systems, Singapore
Management University. \protect\\
E-mail: \{zyang, jiekeshi, bowenxu, davidlo\}@smu.edu.sg, xinzhou.2020@phdcs.smu.edu.sg
\IEEEcompsocthanksitem H.M. Asyrofi is with PropertyGuru Group. E-mail: mhilmiasyrofi@gmail.com.
\IEEEcompsocthanksitem D. Han is with Royal Holloway, University of London. E-mail: DongGyun.Han@rhul.ac.uk.
}
\thanks{Manuscript received April 19, 2005; revised August 26, 2015.}}

%
%

\markboth{Journal of IEEE Transactions on Software Engineering,~Vol.~14, No.~8, September~2022}%
{Shell \MakeLowercase{\textit{et al.}}: Bare Demo of IEEEtran.cls for Computer Society Journals}
\IEEEtitleabstractindextext{%
\begin{abstract}
  With the wide adoption of automated speech recognition (ASR) systems, it is increasingly important to test and improve ASR systems.
  However, collecting and executing speech test cases is usually expensive and time-consuming, motivating us to strategically prioritize speech test cases.
  A key question is: \emph{how to determine the ideal order of collecting and executing speech test cases to uncover more errors as early as possible?}
  Each speech test case consists of a piece of \textit{audio} and the corresponding \textit{reference text}.
  In this work, we propose \toolname (\textbf{PR}i\textbf{O}ritizing s\textbf{P}eec\textbf{H} t\textbf{E}s\textbf{T}), a tool that predicts potential error-uncovering speech test cases only based on their reference texts.
  Thus, \toolname analyzes test cases and prioritize them without running the ASR system, which can analyze speech test cases at large scale.
  We evaluate 6 different prioritization methods on 3 ASR systems and 12 datasets.
  Given the same testing budget, we find that our approach uncovers $12.63\%$ more wrongly recognized words than the state-of-the-art method.
  We select test cases from the prioritized list to fine-tune ASR systems and analyze how our approach can improve the ASR system performance. 
  Statistical tests show that our proposed method can bring significantly larger performance improvement to ASR systems than the existing baseline methods.
  Furthermore, we perform correlation analysis and confirm that fine-tuning an ASR system using a dataset, on which the model performs worse, tends to improve the performance more.
\end{abstract}

}

\maketitle

\section{Introduction}
\label{sec:intro}

The Automated Speech Recognition (ASR) technique has been widely adopted in the interaction between humans and intelligent systems, including virtual assistants~\cite{8301638}, robot control~\cite{robot_control}, and smart healthcare~\cite{healthcare}. 
For example, the Voice Control\footnote{\url{https://support.apple.com/en-us/HT210417}} supported by ASR systems enables users to interact with apps and operating systems with voice commands. 
In addition, modern video platforms (e.g., YouTube) use speech recognition to automatically generate captions to make video content more accessible. 
However, low-quality ASR systems can cause negative impacts on user experience. As shown in recent studies~\cite{strayer2014measuring,cooper2014mental}, vehicle systems that fail to understand spoken commands can distract to drivers, which may cause fatal accidents. 
Ramesh et al.~\cite{bitch} found that well-known ASR systems produce text content highly inappropriate for kids while transcribing YouTube Kids' videos. 
Besides, the operational environments of ASR systems are continuously changing. ASR systems are applied in multiple regions where people have different accents, and new words emerge (e.g., COVID-19, which has not been widely used until recently).

Consequently, ASR system testing~\cite{crossasr,crossasrpp,aequevox,ASRTest} is a vital task for which developers must exert great effort.
Consider the following scenarios: Amy and Bob, as ASR system testers, are responsible for collecting and executing speech test cases that can uncover as many errors as possible in their target ASR systems. 
This goal is typically quantified by counting the number of wrongly transcribed words or characters. 
These test cases that expose the weakness of ASR systems are used to further improve the quality of the ASR systems by fine-tuning them. 
Amy and Bob are facing different scenarios. 

\vspace{0.2cm} 
\noindent \textbf{Scenario 1.} 
The company that Amy works for plans to extend its ASR services from the US to India and UK, where users have different accents. 
A large corpus of texts is available, but without the corresponding accented audio.
To adapt the existing ASR systems to the new regions, Amy is asked to collect some additional audio from the local people to uncover errors in the systems. 
But due to the budget constraints, Amy can only select a limited number of texts that will most likely be wrongly recognized by the ASR system and collect the corresponding audio.

\vspace{0.2cm} 
\noindent \textbf{Scenario 2.} 
To provide better services, Bob's company tries to implement a new ASR system using the latest deep neural network architecture.
Although there are already a large number of available speech test cases, the task is practically constrained by insufficient time and resources to execute all of them. 
Bob needs to prioritize the test cases to be executed to uncover more errors early.

\vspace{0.2cm} 
A speech test case $\langle u, r\rangle$ contains a piece of audio $u$ and the corresponding ground-truth reference $r$. 
The reference $r$ is a text consisting of a sequence of words. In Scenario 1, Amy only has a corpus of references and needs to decide which references are more valuable to annotate and create test cases. In Scenario 2, Bob already has enough speech test cases but needs to decide which subset to execute with higher priority.
Issues in both scenarios can be mitigated by test case prioritization~\cite
{testsurvey}: 
\begin{quote}
    {\em How to decide the ideal order of collecting and executing speech test cases so that more errors can be uncovered even if the testing is prematurely terminated?}
\end{quote}
Arbitrarily collecting and executing test cases may not lead to good testing results. 
Thus, a method to prioritize speech test cases is desired. 
With the help of such a method, testers can strategically collect and execute test cases in a more efficient way.

To achieve the goal, we propose \toolnamenospace, a tool that can effectively prioritize speech test cases.
Compared with previous methods~\cite{crossasr,icassp2021,phonme_rich}, \toolname has the following advantages.
First, \toolname conducts a fine-grained (i.e., word level) analysis to compare test cases, while the previous method~\cite{crossasr} treats all the failed test cases equally even when their ability in uncovering errors is different.
Second, the word-level analysis allows us to leverage the knowledge learned by pre-trained language models (e.g., BERT~\cite{bert} and RoBERTa~\cite{RoBERTa}) to accurately predict the words in a reference text that are likely to be wrongly recognized by an ASR system. However, previous methods~\cite{icassp2021,phonme_rich} based on phonemes have to build a small predictor from scratch as no such pre-trained models are available for phonemes.
Third, rather than simply taking the model output to prioritize test cases, we design a more comprehensive \emph{error score} to estimate the ability of a test case to reveal errors.
A higher error score indicates that a test case tends to uncover more errors and is more worth executing. 
The test cases are prioritized based on the error score of each test case.

To evaluate \toolnamenospace, we select 6 popular speech test case prioritization methods. 
Shinohara~\cite{phonme_rich} proposes to select phonetically rich test cases (i.e., the phonemes should follow a uniform distribution). 
Asyrofi et al.~\cite{crossasr} use an error predictor to predict whether a speech test case is a failure. 
Awasthi et al.~\cite{icassp2021} use an error predictor that can predict which phonemes are likely to be wrongly recognized (i.e., the phoneme error predictor). They also combine phoneme diversity with the phoneme error predictor to analyze test cases. We consider two variants of Awasthi et al.'s method~\cite{icassp2021}: the phoneme error predictors that consider and do not consider phoneme diversity.
We take the phoneme error predictor from Awasthi et al.'s work~\cite{icassp2021} and use it as another method.
Besides, we also include random selection (randomly sampling a certain amount of test cases constrained by the testing budget) as it is the most commonly considered baseline method.
The cost of collecting and executing speech test cases mainly depends on the duration of the audio.
In our experiment, the testing budget is the duration of the audio in a test suite to be executed.

We experiment with 3 ASR systems: Quartznet~\cite{quartznet}, HuBERT~\cite{HuBERT}, and wav2vec 2.0~\cite{wav2vec2}.
Quartznet, as a representative of small-scale and supervised-learning ASR models, is used by Awasthi et al.~\cite{icassp2021} to evaluate their phoneme error predictor. 
We additionally consider two Transformer-based~\cite{vaswani2017attention} representation learning models that achieve state-of-the-art performance.
In total, we experiment with 6 prioritization methods and evaluate them on 12 datasets and 3 ASR systems. 
We apply different methods to prioritize the same group of test cases.
Then, we execute test cases (using the same testing budget) from each ranked list and compare the errors uncovered by each method.
To mitigate the effects of randomness, we run the experiments with different configurations three times and compare the average values.
Based on the large-scale experiments, we compare the investigated methods from multiple perspectives.

We find that \toolname can better prioritize test cases to uncover more errors. 
More specifically, out of a total of 36 cases (3 ASR systems $\times$ 12 datasets), \toolname outperforms the other baseline methods in 35 cases.
Given the same testing budget, our method uncovers $12.63\%$ more word errors than the second-best method on average.
We also evaluate the values of using the prioritized test cases to improve ASR system performance. 
We select the same number of test cases from the test suites prioritized by different methods and call them the {\em fine-tuning sets}, which are used to fine-tune ASR systems.
We conduct statistical tests to compare the changes in model performance brought by different prioritization methods. 
The results demonstrate that our method can provide significantly larger model performance improvement than other investigated methods. 

A correlation analysis is conducted to further understand what kind of fine-tuning sets can better improve ASR performance.
We consider the 3 features of a fine-tuning set: 
the original model (i.e., an ASR system before fine-tuning) performance on this dataset, and two metrics describing how the phonemes distribute in this fine-tuning set. 
The latter two metrics are usually used to measure the diversity of an ASR dataset~\cite{7178848}.
The result suggests that the original model performance on the fine-tuning set shows a stronger correlation than the features related to phoneme diversity. 
It contradicts the previous belief that phoneme diversity can help select representative subset of test cases to improve ASR systems~\cite{DBLP:journals/corr/abs-2203-09829,icassp2021}.

The contributions of this paper include:
\begin{itemize}[leftmargin=*]
    \item We propose \toolnamenospace, a tool that is designed to predict word errors in ASR systems. Based on \toolnamenospace, we design an algorithm that can effectively prioritize test cases for testing and improving ASR systems. 
    \item We conduct large-scale experiments (obtaining \numbermodel ASR systems) to analyze the effectiveness of speech test case prioritization methods. The results show that our proposed method can achieve state-of-the-art results in both uncovering more errors. 
    \item We find that our method also achieve state-of-the-art results in improving the model performance. Our analysis confirms that fine-tuning an ASR system on a dataset on which it produces more errors leads to higher model performance improvement.
\end{itemize}

The rest of this paper is organized as follows. Section~\ref{sec:motivation} formalizes the speech test case prioritization problem and discusses the relevant baseline methods. In Section~\ref{sec:methodology}, we describe the details of our proposed approach. Section~\ref{sec:experiment} explains the experiment settings.
Section~\ref{sec:result} reports research questions and experiment results. We discuss the threats to validity and related work in Section~\ref{sec:discussion} and Section~\ref{sec:related_work}, respectively. Finally, we conclude the paper and present future work in Section~\ref{sec:conclusion}.

\section{Background}
\label{sec:motivation}
This section formalizes the speech test case prioritization task and describes four related methods.

\subsection{Speech Test Case Prioritization}

The key question in both two testing scenarios presented previously is: {\em how to decide the ideal order of collecting and executing speech test cases so that more errors can be uncovered even if the testing is prematurely terminated?}
In this paper, we assume that the prioritization process is based on the reference texts in the speech test cases.
Following the definition of test case prioritization from previous works, e.g.~\cite{testsurvey}, we formulate the \textit{speech test case prioritization} as follows. 
We denote an existing ASR system under test as $\mathcal{A}$. 
The ASR system takes as input a piece of audio $u$ and produces a transcript $r$, which is represented as $\mathcal{A}: u \mapsto r$. 
A speech test suite $\mathcal{T} = \{(u_1, r_1), (u_2, r_2), \cdots, (u_n, r_n)\}$ is given.
The set of permutations of $\mathcal{T}$ is denoted as $\mathcal{P}(\mathcal{T})$.
Our goal is to find an optimal permutation $\mathcal{T}^{*} \in \mathcal{P}(\mathcal{T})$ such that:
\begin{equation}
    f(\mathcal{T}^{*}, k) \geq f(\mathcal{T}', k), \forall \mathcal{T}' \in \mathcal{P}(\mathcal{T})
\end{equation}
In the above equation, the function $f(\mathcal{T}^{*}, k)$ quantifies the errors uncovered when top $k$ test cases are executed in the prioritized test suite $\mathcal{T}^{*}$.
The commonly used metrics are the \emph{word error rate} ($\mathit{WER}$) and \emph{character error rate} ($\mathit{CER}$), which we will explain in Section~\ref{subsec:evaluation_metrics}.


\subsection{Baseline Methods}
\label{subsec:baseline}


The test case prioritization methods~\cite{919106,10.1145/3460319.3464810,10.1145/3236024.3236053} designed for conventional software systems are not applicable to ASR systems as the concepts used (e.g., branch coverage metrics) are not well-defined for ASR systems.
Recently, a series of methods~\cite{surprise,DeepHunter,DeepXplore,sensei} have been proposed for selecting test cases for other DNN models (e.g., computer vision and natural language process systems). 
Ma et al.~\cite{ma_tosem} summarized these methods and conducted an empirical study to evaluate their effectiveness. 
These methods work by analyzing the inputs (i.e., the audio in our case) and the outputs from models under test.
In ASR testing, however, the audio is not always available. 
As a result, the metrics that require executing the deep learning models (e.g., neuron coverage~\cite{harel2020neuron}, surprise adequacy~\cite{surprise}, and model uncertainty~\cite{DeepGini,ASRTest}) are not applicable in ASR testing. 
We consider a realistic ASR testing scenario and only utilize the reference text to prioritize speech test cases.
We present methods (from both software engineering and speech recognition communities) that are designed to prioritize speech test cases.

\begin{table}[!t]
    \caption{The phonemes and triphones of the word `speech.'}
    \begin{tabular}{ccc}
    \hline
    Word   & Phonemes         & Triphones                         \\ \hline
    speech & {[} s, p, i, \ipa{{tS}} {]} & {[}$\langle$\#-s-p$\rangle$, $\langle$s-p-i$\rangle$, $\langle$p-i-\ipa{{tS}}$\rangle$, $\langle$i-\ipa{{tS}}-\#$\rangle${]} \\ \hline
    \end{tabular}
    \label{tab:phonemes}
    \end{table}


\subsubsection{Phoneme-Rich Selection}



As shown in Table~\ref{tab:phonemes}, an English word can be decomposed into a list of phonemes. For example, the word `speech' can be decomposed into the phonemes `s', `p', `i', and `\ipa{{tS}}'. A triphone, also known as the context-dependent phoneme, is a triple consisting of three phonemes, which captures the context of a phoneme. 
Shinohara~\cite{phonme_rich} suggests sampling a test suite that follows a desired distribution over phonemes or triphones. 
The intuition is that texts in a corpus are usually distributed not in a uniform way (i.e., frequent phonemes might appear much more times than less frequent ones). 
As a result, random sampling will make the selected data still follow a similar distribution.
When using the randomly sampled data to train an ASR system, the system tends to perform poorly on the less frequently appearing phonemes.
Shinohara~\cite{phonme_rich} proposed a Phoneme-Rich method to favor texts that appear less frequently. This method defines a utility function of a test suite $S$:
\begin{equation}
    J(S) = \sum_{i=1}^{|P|} \pi_{i} \log f_i(S)
    \label{eq:utility}
\end{equation}
$|P|$ represents the total number of phonemes, $\pi_i$ is the desired distribution of the $i^{th}$ phoneme (e.g., uniform distribution in~\cite{phonme_rich}), and $f_i(S)$ is the number of times that the $i^{th}$ phoneme appears in $S$. This function is proved to have the following {\em submodularity}~\cite{phonme_rich} feature. Consider two texts $s_1$ and $s_2$; for simplicity, we assume that they both have one phoneme. The text $s_1$ has the phoneme that is less frequently appearing than $s_2$ in $S$. Adding $s_1$ into $S$ can produce a larger utility gain than adding $s_2$ into $S$, that is, $J(S \cup \{s_1\}) - J(S) > J(S \cup \{s_2\}) - J(S)$. It means that this function will favor the texts containing less frequent phonemes, and selecting these sentences can lead to the goal of having a test suite of uniform distribution. 
For a group of test cases to be evaluated, the proposed method computes utility gain of each test cases, i.e., $J(S \cup \{s\})  - J(S)$. 
The test case with the highest gain is augmented into $S$. This greedy selection process produces a prioritized list of given speech test cases based on their reference texts.




\subsubsection{Phoneme Error Predictor}
\label{subsubsec:phoneme-error}
Awasthi et al.~\cite{icassp2021} utilize a few outputs from an ASR system to train an error predictor. 
Awasthi et al.~\cite{icassp2021} train a 4-layer Bi-LSTM model~\cite{hochreiter1997long} that can predict error at the phoneme level (hereafter, we call it Phoneme Error Predictor (\emph{PEP})). 
The reference text in a test case is first converted into a list of phonemes as shown in Table~\ref{tab:phonemes}, which is then fed into \emph{PEP} to predict the probability that each phoneme is wrongly recognized. 
Awasthi et al. also consider the phoneme diversity to prioritize test cases using the submodular function value to assign a weight to each phoneme in a text. The intuition is that even a phoneme is highly likely to be wrongly recognized; we should avoid over-selecting it if this phoneme has appeared many times in the selected texts. 
Awasthi et al.~\cite{icassp2021} combine the submodular function and the results from \emph{PEP} to compute a metric to evaluate the value of a text and give higher priority to the ones with higher values. The metric is formulated as follows:
\begin{equation}
    \frac{1}{n} \sum_{\varphi \in \mathcal{P}} c_{\varphi}(S,s)\sum_{j \in \{s_j = \varphi\}}Pr_j(e^j=1|s)
\end{equation}
$Pr_j(e^j=1|s)$ means the probability that an ASR system wrongly predicts the $j$-th phoneme in $s$, $c_{\varphi}(S,s)$ is the value gain of the submodular function if $s$ is selected into $S$, and $n$ is the number of phonemes in the text $s$. We call this method that considers both phoneme richness and errors as \emph{PEP-D}.

\subsubsection{Sentence Error Predictor}
\label{subsubsec:sentence-error}


Asyrofi et al.~\cite{crossasr} design \emph{CrossASR}, a differential testing framework that leverages synthetic audio to test ASR systems. 
\emph{CrossASR} can access two types of ASR systems: one ASR system under test and several ASR systems for cross-reference. 
\emph{CrossASR} generates a speech test case by synthesizing audio from a text.
A test case is called a \emph{failure} only when the system under test fails to recognize the audio and at least one system for reference can recognize the audio correctly.
Test case prioritization is one of the core elements for this framework to work effectively. 
\emph{CrossASR} collects error information of the system under test and trains a failed test case predictor during the executing time, which takes as input a text and produces a binary label indicating whether the corresponding audio will be correctly recognized or not. 
This predictor allows \emph{CrossASR} to strategically prioritize the texts to synthesize audio.

For a group of texts to be prioritized, \emph{CrossASR} first leverages this predictor to estimate the probability that a text can lead to a failure. 
Then it ranks the texts according to their probabilities in descending order. 
\emph{CrossASR} synthesizes test cases for prioritized texts until the budget is fully used, that is, reaching the execution time limit. 
Although \emph{CrossASR} is designed as a differential testing tool, its failed test case predictor serves the purpose of test case prioritization, so we include it as a baseline method to compare. 

\section{Methodology}
\label{sec:methodology}

This section describes our proposed method, \toolname (\textbf{PR}i\textbf{O}ritizing s\textbf{P}eec\textbf{H} t\textbf{E}s\textbf{T}).
\toolname addresses the limitations in the existing methods to prioritize test cases:
\begin{enumerate}[leftmargin=*]
  \item \emph{CrossASR}~\cite{crossasr} predicts errors at the sentence level, i.e., two failed test cases (e.g., one uncovering 1 word errors and the other uncovering 10 word errors) are treated equally.
  ASR developers care about the number of word errors.
  To overcome this limitation, we propose to tackle the task of predicting error the word-level error prediction (Section~\ref{subsec:localization}).
  \item Awasthi et al.~\cite{icassp2021} aim to predict errors at the phoneme level. However, they only train a simple model (i.e., a 4-layer LSTM model) on a small dataset, which produces an ineffective error predictor. 
  The word-level error prediction task allows us to leverage knowledge in existing powerful pre-trained language models: producing contextualized embeddings for each word in a reference text and using the embeddings to predict word errors (Section~\ref{subsec:error_predictor}).
  \item Rather than simply taking the model output to prioritize test cases, we design a more comprehensive metric to estimate the ability of a test case to reveal errors (Section~\ref{subsec:selection}).
\end{enumerate}

\subsection{Word-level Error Prediction}
\label{subsec:localization}

To efficiently test and improve ASR systems, developers collect and execute the test cases that trigger more errors, i.e., wrongly recognized words.
However, collecting data blindly may not produce good testing results. 
Thus, we provide a proxy to estimate the probability of each word in a text being wrongly recognized by an ASR system and then use the proxy to prioritize speech test cases. 
We call the task of predicting errors at the word level as \emph{word-level error prediction}, which performs low-granularity analysis of the ASR system performance. 
We define the task as follows.
Assuming that we have an ASR system under test denoted by $\mathcal{A}$, given a piece of audio $u$ whose corresponding ground truth reference is $r$, the ASR system transcribes the audio $u$ into $\hat{r}$. 
The pair $\langle u, r\rangle$ is a {\em failed test case} for $\mathcal{A}$ if $r$ and $\hat{r}$ are not exactly the same. 
To quantify the errors, we measure the difference between $r$ and $\hat{r}$ by first converting $r$ and $\hat{r}$ into two lists of words (separated by whitespace) and then using the Levenshtein distance~\cite{10.1145/375360.375365} to capture the difference.

\begin{figure}[t!]
    \centering
	\includegraphics[width=1\linewidth]{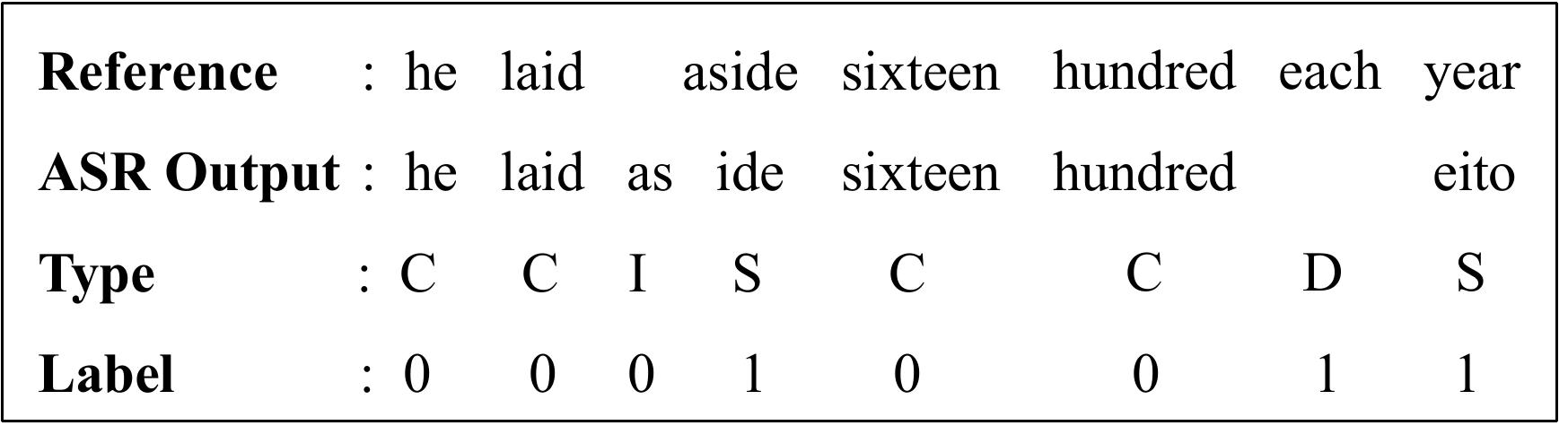}
    \caption{An example of the reference (i.e., the ground truth output for a test case) and the output from an ASR system). {\tt Type} caputres the edit distance between the reference and ASR output. All the `{\tt S}' and {\tt D} edits are treated as errors.}
    \label{fig:token-classification}
\end{figure}

Figure~\ref{fig:token-classification} illustrates an example of the difference between a reference and an ASR output by using the {\tt power-asr} library.\footnote{\url{https://github.com/NickRuiz/power-asr}}
In the figure, {\tt Type} means the types of difference, and {\tt Label} indicates whether a word in the reference is wrongly recognized (1 means wrong transcription and 0 otherwise). 
There are four types of differences in Figure~\ref{fig:token-classification}. 
`{\tt C}' denotes a correct match, i.e., the word in the reference is the same as the word in the ASR output. `{\tt I}' denotes an insertion, i.e., the word in the ASR output is not in the reference. `{\tt D}' means deletions, i.e., a word in the reference does not appear at the same position in the ASR output. `{\tt S}'  represents substitutions, i.e., a word in the reference is recognized as another word. 
In our study, we only predict word errors caused by `{\tt S}' and `{\tt D}' edits, as the original word is either replaced or deleted. 
This task is to predict the words in the reference texts that are wrongly recognized. 

Based on the edit distance between the reference and the ASR output, we derive a label for each word in the reference standing for whether it is a word error. 
We aim to predict the label for each word solely based on the reference $r$ without accessing the corresponding audio and running the ASR system. 
This paper considers the text-based task for two reasons.
On the one side, processing audio is usually time-consuming while it is more efficient to analyze the texts, allowing test case prioritization at scale. 
On the other side, the audio may be unavailable and texts need to be prioritized to collect audio efficiently (e.g., the testing scenario 1 in Section~\ref{sec:intro}).

\subsection{\toolname}
\label{subsec:error_predictor}

The word-level error prediction task can be viewed as a sequence labeling problem on texts: given a reference text consisting of several words, assigning a label to each word to indicate whether an ASR system will wrongly recognize it. 
However, building a model to tackle this task is non-trivial, especially considering that there is not much data available to train the model.
Recently, large pre-trained language models~\cite{bert,RoBERTa,9240704} have demonstrated excellent performance in the field of natural language processing and benefited a series of text processing tasks in particular~\cite{mengge-etal-2020-coarse}. 
These pre-trained models also enable us to train a model with a limited amount of data.
Such pre-trained language models have achieved success in facilitating software engineering activities~\cite{9825884,9796213,10.1145/3551349.3560421,10.1145/3551349.3556964}.
We leverage BERT~\cite{bert} (one of the most famous models) to build a sequence labeling model to predict word errors for an ASR system without running it. 
We now introduce input preparation and the feedforward process, how to collect training data from the ASR system behaviors, and the loss function to fine-tune the model.

\subsubsection{Model Architecture}
\begin{figure}[t!]
	\centering
	\includegraphics[width=1\linewidth]{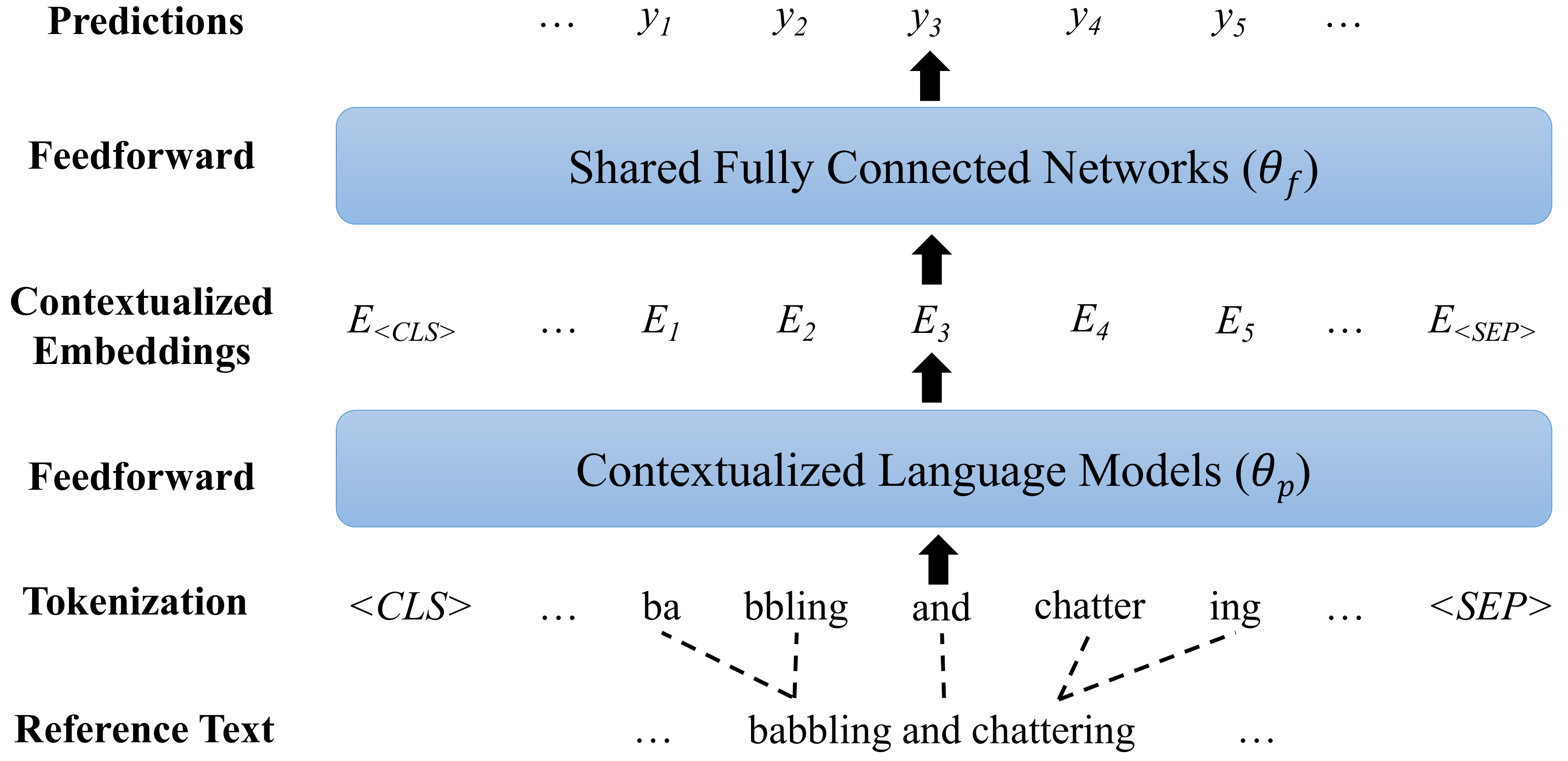}
	\caption{The architecture of \toolnamenospace. After obtaining the contextualized embeddings from BERT, we use a shared fully connected network to predict the label for each token in the input sequence. 
  }
	\label{fig:model}
\end{figure}

To alleviate the out-of-vocabulary problem~\cite{shi2022identifier}, BERT adopts the WordPiece tokenizer~\cite{wordpiece} to split a word that is not in its vocabulary into several {\em subwords}. 
This process is called \emph{tokenization}.
Taking `{\tt babbling}' and `{\tt chattering}' as examples, the WordPiece tokenizer will convert them into $\langle$`{\tt ba}', `{\tt bbling}'$\rangle$ and $\langle$`{\tt chatter}', `{\tt ing}'$\rangle$, respectively. 
Please note that a reference text of a speech test case consists of a list of \emph{words}. By splitting a word into \emph{subwords}, the text is tokenized into an \emph{input sequence} made up of \emph{tokens} of \emph{subwords}. 
As shown in Figure~\ref{fig:model}, BERT puts two special tokens (i.e.,$\langle CLS \rangle$ and $\langle SEP \rangle$) at the beginning and the end of the tokenized input sequence. 
Then, BERT produces a contextualized embedding for each token in the input sequence. 
We use a fully connected network to process the embeddings and obtain the predicted label for each token. Note that the fully connected network is shared for processing each embedding.

\subsubsection{Data Preparation}
To fine-tune this model, we need the ground truth labels for each token. 
We use the label 1 to represent that a token is wrongly recognized and 0 otherwise.
\toolname requires a small amount of labeled speech test cases and their corresponding outputs from an ASR to observe the behaviors of the ASR system under test.
Note that the baselines~\cite{icassp2021,crossasr} also adopt the same setting.
The label preparation process for the tokens (i.e., subwords) is as follows. 
We keep the labels for subwords the \emph{same as the label for the original word that they are tokenized from}. 
For example, if the word `{\tt babbling}' is a word error, its two subwords `{\tt ba}' and `{\tt bbling}' shall be labeled as errors (i.e., have a label of $1$) as well. 
The beginning and the end of a BERT input are two special tokens, which are assigned with the label -100 (the default value to let PyTorch ignore this label when computing the loss).
By doing so, we obtain an aligned label sequence sharing the same length as the model input sequence. 

\subsubsection{Loss Function}
We denote \toolname as $f$, which is parameterized by $\theta_p$ and $\theta_f$, the parameters for the BERT model and the fully connected network, respectively. 
Assuming that $y_i$ is the ground truth label for a token $x_i$ in the input and $f(x_i)$ is the prediction on this token, we fine-tune the model by optimizing the objective:
\begin{equation}
  \arg \min_{\theta_p, \theta_f} \sum_{x \in D} \sum_{i=1}^{n} -y_{i} \log (f(x_i))
\end{equation}
In the above equation, $n$ is the number of tokens in an input sequence $x$, and $D$ is the dataset for training this model. 
Note that the gradient can flow through the parameters of both the BERT model and the fully connected network. 
To minimize the loss function on the training data, we use back-propagation to update both parameters $\theta_p$ and $\theta_f$ when fine-tuning ASR systems.

\subsection{Prioritize Test Cases using \toolnamenospace}
\label{subsec:selection}

The test cases are prioritized based on their likelihood of uncovering errors estimated by the approaches.
Imagine that there are two failed test cases: the one uncovering only one word error and the other uncovering 10 word errors. 
Intuitively, the second one would be preferred as it uncovers more errors.
Following this intuition, our prioritization method values test cases that can uncover more word errors.
\emph{CrossASR}~\cite{crossasr} predicts errors at the sentence level (i.e., the probability that the transcription is not fully correct). 
However, a failed test case can uncover multiple (sub)word errors.
To overcome the limitation of CrossASR~\cite{crossasr}, we design our approach to be capable of perceiving errors at the word-level. To achieve the goal, we define the {\em error score} as the average probability of each subword being an error in the input sequence, which approximates the `\textit{error density}' in a test case. 
Given a reference text $r$, we use $P(r_i=1|r)$ to represent the probability that the $i^{th}$ subword in $r$ is an error. 
Assuming that there are $n$ subwords in $r$, the error score is formally defined as:
\begin{equation}
  error\_score(r) = \frac{1}{n} \sum_{i=1}^{n} P(r_i=1|r)
  \label{eq:error_score}
\end{equation}
Last, our approach produces the error score of each of the given references and sort them accordingly for the prioritization purpose.

\section{Experiment Design and Settings}
\label{sec:experiment}
This section explains the settings of our experiments, including the investigated ASR systems and datasets, implementation details, and the metrics to evaluate the ASR system performance.

\begin{figure}[t!]
  \centering
\includegraphics[width=1\linewidth]{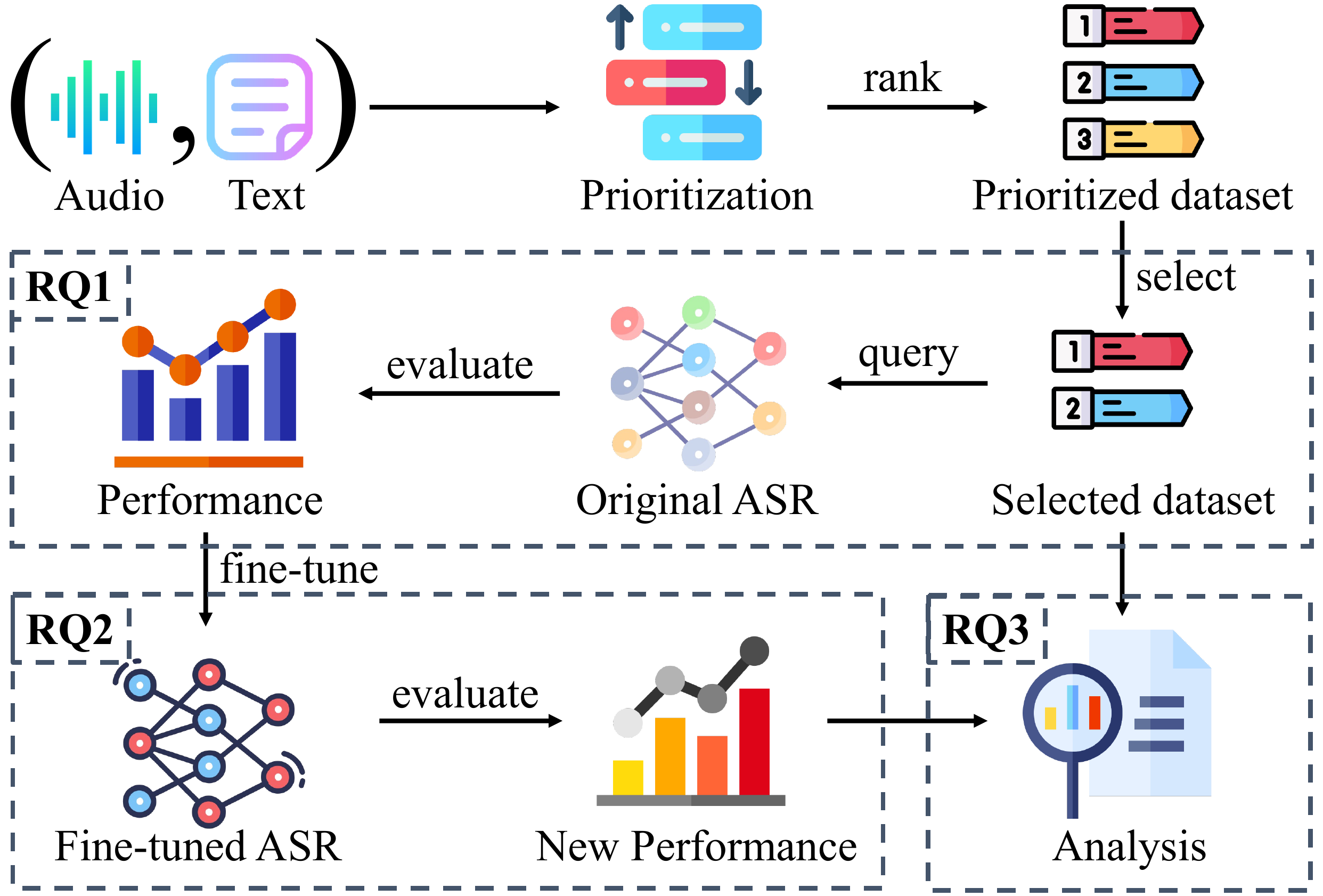}
  \caption{An overview of evaluating and improving ASR systems on prioritized test cases and relationships between three investigated research questions.}
  \label{fig:overview}
\end{figure}

\subsection{Overview}
\label{subsec:overview}

Figure~\ref{fig:overview} provides an overview of how we systematically compare the speech test case prioritization methods. 
Given a corpus of speech test cases, we apply a method to rank all the test cases.
Depending on the testing budget, the tester selects some test cases from the ranked list to produce the {\em selected dataset}.  
We evaluate the original model on the selected dataset to collect their performance metrics (explained in Section~\ref{subsec:evaluation_metrics}).
We fine-tune ASR systems using the selected dataset to obtain new models. 
We compare the performance of the original and new models on an unseen test set (mutually exclusive to datasets for training and fine-tuning the ASR system) to analyze which method is more effective in finding the test cases that can better improve ASR systems.

\subsection{ASR systems and Datasets}
\label{subsec:models_datasets}


Our study considers three ASR systems: Quartznet~\cite{quartznet}, HuBERT~\cite{HuBERT}, and wav2vec 2.0~\cite{wav2vec2}.
Quartznet~\cite{quartznet} is an end-to-end model that uses multiple blocks with residual connections. 
Quartznet is trained in a supervised way and can achieve near state-of-the-art performance with significantly fewer parameters.
We include Quartznet and treat it as a representative of small-scale and supervised-learning models. 
Quartznet is also used by Awasthi et al.~\cite{icassp2021} to evaluate their phoneme error predictor. 
We consider two Transformer-based representation learning models that adopt unsupervised learning.
The wav2vec 2.0~\cite{wav2vec2} model achieved state-of-the-art performance when it was proposed.
Hsu et al. proposed HuBERT~\cite{HuBERT}, which learns by predicting the masked hidden units. 
HuBERT can either match or even improve upon the performance of wav2vec 2.0.

The ASR systems are trained on popular datasets including Librispeech~\cite{librispeech}, Common Voice~\cite{common-voice} and Libri-Light~\cite{Libri-Light}.
It is inappropriate to evaluate the test case prioritization methods on these datasets as they shall be already seen by the models and may introduce unexpected bias into the experiment.
We follow the work by Awasthi et al.~\cite{icassp2021} who used a subset of two datasets: IndicTTS~\cite{IndicTTS} and L2-Arctic~\cite{l2arctic}. 
As the IndicTTS is not publicly available, to ensure that the following researchers can replicate and extend our work easily, we use 12 datasets from L2-Arctic~\cite{l2arctic}; each dataset contains 1,130 audio files on average.
We randomly sample 10\% of the audio files from each dataset and use them as the test set. 
Table~\ref{tab:statistics} describes the performance of three ASR systems on sampled test set of these 12 datasets.
The first column contains the names of each sub-dataset.

\begin{table}[!t]
  \centering
  \caption{Statistics of models under test and their performance on different datasets.}
  \begin{tabular}{ccccccc}
  \hline
  \multirow{2}{*}{Datasets} & \multicolumn{2}{c}{QuartzNet~\cite{quartznet}} & \multicolumn{2}{c}{HuBERT~\cite{HuBERT}} & \multicolumn{2}{c}{wav2vec2.0~\cite{wav2vec2}} \\ \cline{2-7} 
                           & $\mathit{WER}$           & $\mathit{CER}$           & $\mathit{WER}$          & $\mathit{CER}$         & $\mathit{WER}$            & $\mathit{CER}$           \\ \hline
  YBAA                     & 20.84         & 7.71          & 11.13        & 3.33        & 17.98          & 6.59          \\
  ZHAA                     & 20.75         & 8.05          & 13.22        & 4.38        & 20.35          & 7.57          \\
  ASI                      & 15.38         & 5.19          & 8.32         & 2.62        & 15.70          & 5.53          \\
  TNI                      & 18.28         & 7.01          & 9.66         & 3.15        & 19.63          & 7.44          \\
  NCC                      & 30.35         & 12.41         & 18.50        & 7.23        & 31.81          & 12.86         \\
  TXHC                     & 22.79         & 9.10          & 12.31        & 4.50        & 21.97          & 8.42          \\
  EBVS                     & 36.49         & 17.24         & 24.29        & 10.76       & 37.22          & 17.02         \\
  ERMS                     & 25.94         & 9.85          & 14.24        & 5.01        & 26.45          & 9.53          \\
  YDCK                     & 16.84         & 6.40          & 11.33        & 3.98        & 18.50          & 7.25          \\
  WKWK                     & 20.60         & 8.10          & 12.17        & 4.16        & 18.31          & 6.82          \\
  THV                      & 39.57         & 17.77         & 28.79        & 12.59       & 39.78          & 17.97         \\
  TLV                      & 46.39         & 20.81         & 35.91        & 16.20       & 45.57          & 20.22         \\ \hline
  \end{tabular}
  \label{tab:statistics}
  \end{table}

\subsection{Implementation}
\label{subsec:implementation}
For the implementation of ASR systems, we use the pre-trained ASR systems released by the authors in our experiments.
We use the code of Quartznet~\cite{quartznet} and the pre-trained model from the official repository\footnote{https://github.com/NVIDIA/NeMo} provided by Nvidia. 
The HuBERT and wav2vec 2.0 are both obtained from the huggingface platform.\footnote{https://huggingface.co/} 
For HuBERT, we select a larger version of model, i.e., HuBERT-large.
For the implementation of baseline prioritization methods, we reimplement Shinohara's work~\cite{phonme_rich} on our own as the paper is not accompanied by the corresponding open-source implementations.
We obtain the implementation of \emph{PEP}\footnote{https://github.com/awasthiabhijeet/Error-Driven-ASR-Personalization} and \emph{CrossASR}\footnote{https://github.com/soarsmu/CrossASR} from online repositories released by the authors~\cite{crossasr,icassp2021}. 

We run the experiments with 4 different budget settings, i.e., how many test cases should be executed from the prioritized list. 
We also run each method three times with different random seeds to mitigate the effect of the randomness. 
Given the configuration of 12 datasets, 3 ASR systems, 6 prioritization methods, 4 budgets and 3 random seeds, we obtain \numbermodel models in the experiment. 
Due to the limited space, we describe other hyper-parameter settings (e.g., the learning rate for each model) in an online appendix in the replication package. 
Our experiments were performed on two machines running Ubuntu 18.04 OS with 8 NVIDIA GeForce P100 GPUs on each machine. To facilitate the experiment, we parallelize the fine-tuning process on multiple GPUs.

\begin{figure*}[t!]
  \centering
  \includegraphics[width=4.4 in]{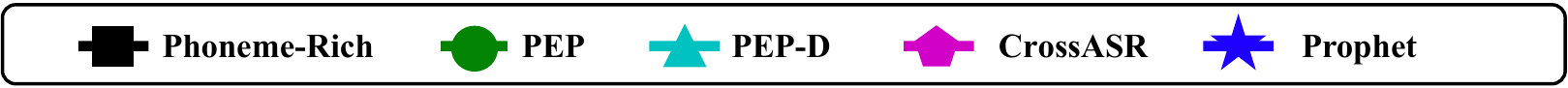}
  \vspace{-0.4cm}
  \end{figure*}
  
  \begin{figure*}[t!]
      \centering
      \begin{subfigure}[c]{0.33\textwidth}
          \includegraphics[height=2.1 in]{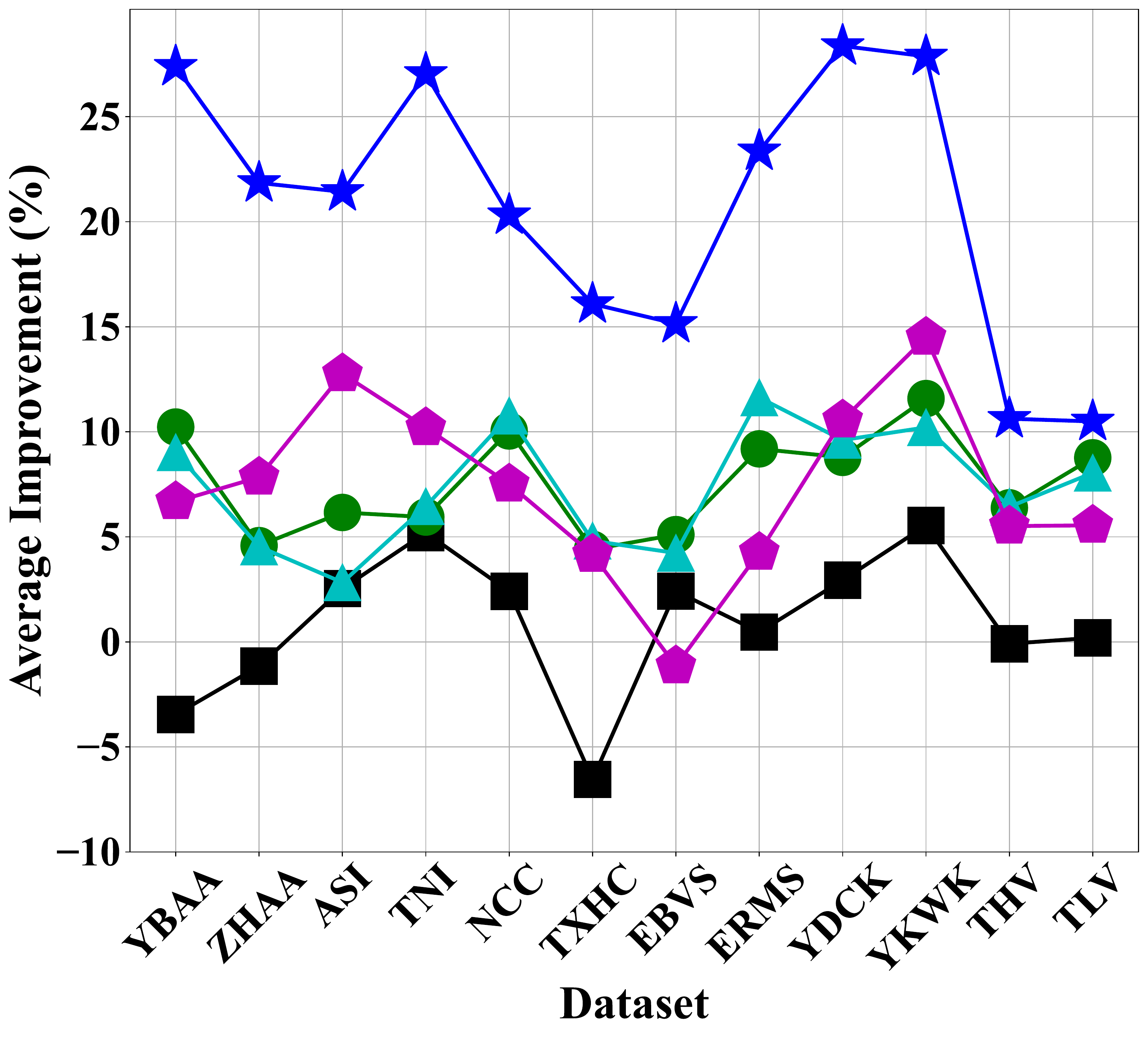}
          \caption{QuartzNet}
      \end{subfigure}
      \quad
      \begin{subfigure}[c]{0.31\textwidth}
          \includegraphics[height=2.1 in]{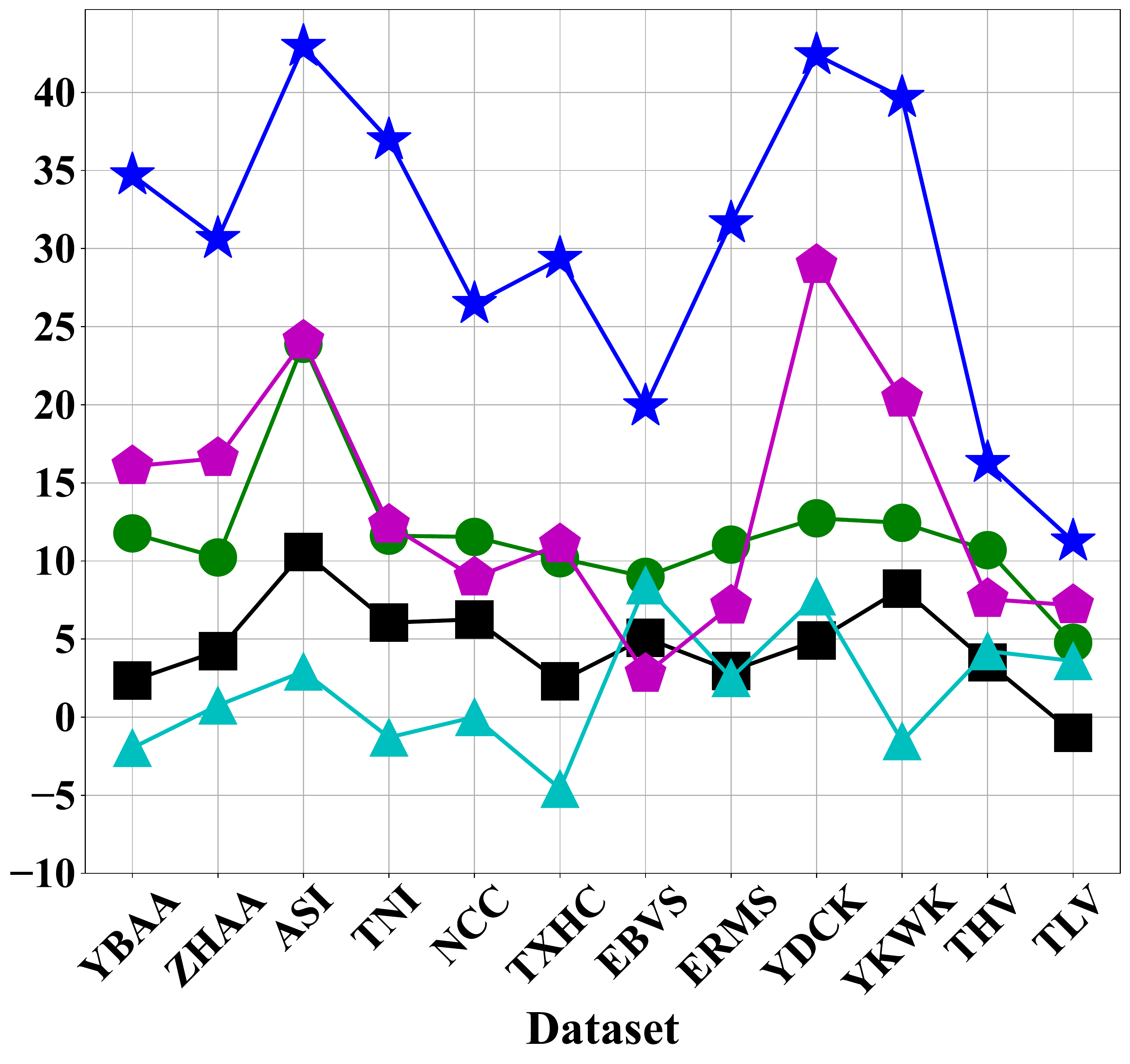}
          \caption{HuBERT}
      \end{subfigure}
      \quad
      \begin{subfigure}[c]{0.31\textwidth}
          \includegraphics[height=2.1 in]{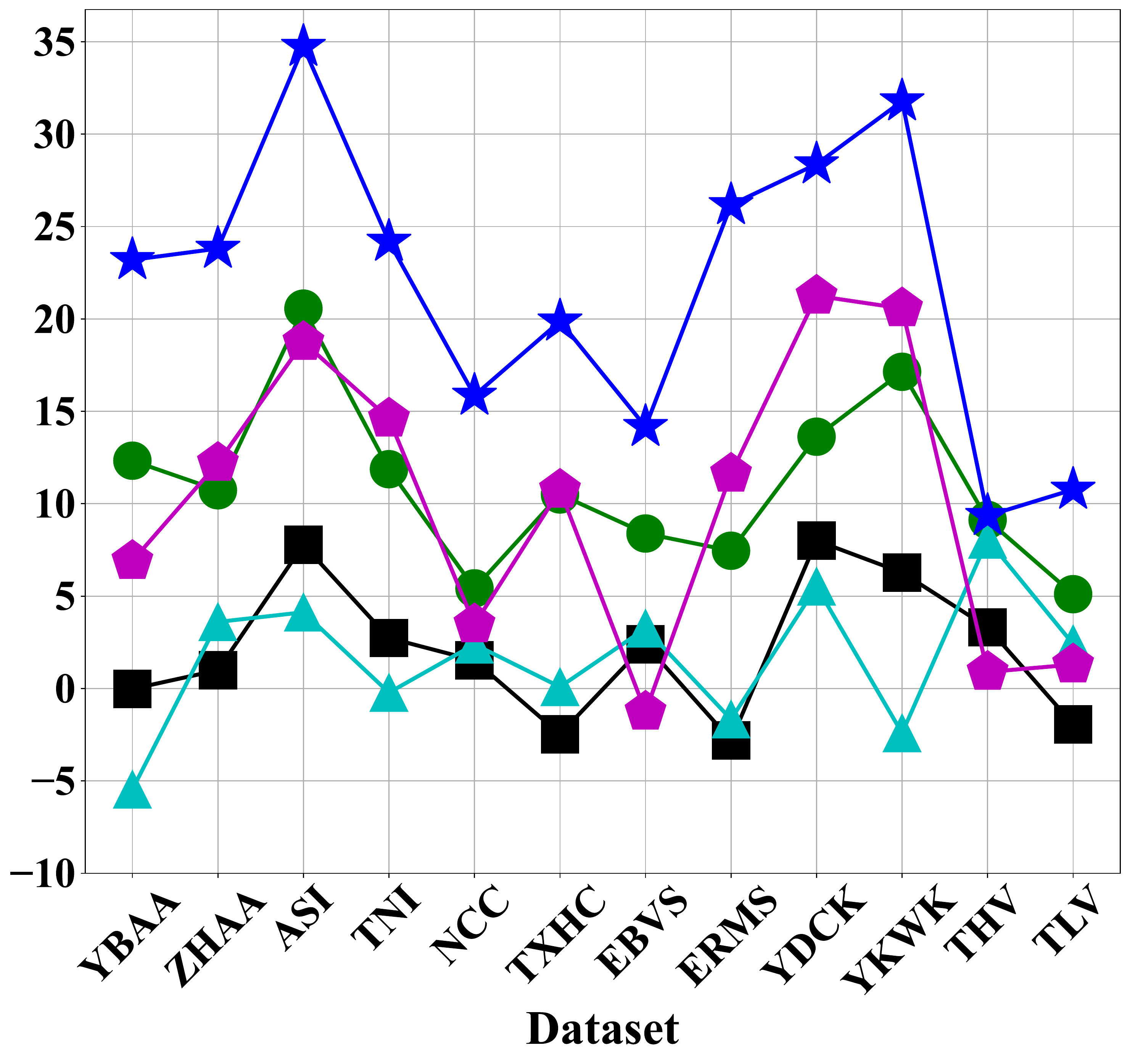}
          \caption{wav2cev 2.0}
      \end{subfigure}
      \caption{The average improvement of different methods against the random selection. The x-axis of each sub-figure represents different datasets, and the y-axis represents the average improvement.}
      \label{fig:RQ1}
  \end{figure*}

\subsection{Evaluation Metrics}
\label{subsec:evaluation_metrics}

The metrics for evaluating test case prioritization algorithms on conventional software systems are not applicable to the ASR testing scenarios.
For example, computing average percentage of fault detection (APFD) requires measuring faults detected by a test suite. 
However, just like other DNN-enabled systems, what constitutes a fault is unclear in ASR testing.
Although Pan et al.~\cite{10.1007/s10664-021-10066-6} mention that APFD can be calculated based on the number of test case failures when the number of faults detected is unavailable, ASR developers prefer more fine-grained error information (e.g., which words or characters are wrongly recognized) rather than just the number of failed test cases. 
As a result, following the practices in evaluating ASR systems, we adopt two commonly used metrics to measure the performance of an ASR system on a certain dataset: {\em word error rate} ($\mathit{WER}$) and {\em character error rate} ($\mathit{CER}$), as the indicators of how a test case prioritization method performs.

As illustrated in Figure~\ref{fig:token-classification}, we can compute the edit distance between the ASR output and the reference text. 
The distance is represented using a list of edit operations. We use $S$, $I$ and $D$ to represent Substitutions, Insertions and Deletions operations, respectively. 
A model's $\mathit{WER}$ on a dataset is the ratio of the three operations in all the words, which can be computed using the following formula:

\begin{equation}
  \mathit{WER} = \frac{\text{Number of ($S$ + $I$ + $D$) edits}}{\text{Total number of words}}
\end{equation}
Similarly, we can also calculate the edit distance at the character level. Having the total number of characters and character-level edits in the dataset, we can compute the $\mathit{CER}$ using the formula:

\begin{equation}
  \mathit{CER} = \frac{\text{Number of ($S$ + $I$ + $D$) edits}}{\text{Total number of characters}}
\end{equation}
A lower $\mathit{WER}$ and $\mathit{CER}$ indicate that an ASR system is better at recognizing the words or characters. 
An effective test case prioritization method should give higher priority to test cases that can uncover more word (character) errors. 
In other words, given the same testing budget, a more effective method should be able to identify a dataset on which an ASR system has a high $\mathit{WER}$ ($\mathit{CER}$).

\section{Results}
\label{sec:result}

This section comprehensively evaluates the proposed \toolname and the baseline methods (Section~\ref{sec:motivation}) on 3 ASR systems and 12 datasets to answer the following three research questions:
\begin{itemize}
  \item RQ1. RQ1. How is the performance of \toolname in uncovering ASR errors?
  \item RQ2. How is the performance of \toolname in enhancing ASR systems performance?
  \item RQ3. What kind of datasets are more effective in improving ASR system performance?
\end{itemize}

\subsection*{RQ1. How is the performance of \toolname in uncovering ASR errors?}

We follow a ratio of 8:1:1 to randomly split each dataset into \emph{selection set}, \emph{validation set}, and \emph{test set}, respectively.
The selection set contains test cases to be prioritized. 
The validation set and test set are used to evaluate the performance of fine-tuned ASR systems.
To train the error predictors (i.e., \toolnamenospace, \emph{CrossASR}~\cite{crossasr} and \emph{PEP}~\cite{icassp2021}), we randomly select 10\% of test cases from the selection set (called \emph{seed data}) and query the ASR system under test.
Then, following the procedure mentioned in Section~\ref{subsec:error_predictor}, we use the outputs from ASR systems to train the error predictors.

Following the previous study~\cite{icassp2021}, to implement the random selection baseline, we randomly choose test cases (using 4 different settings: 50, 100, 200, 400) from the remaining test cases in the selection set. 
By using the {\tt wave} library\footnote{\url{https://docs.python.org/3/library/wave.html}} to measure the total duration of these randomly selected test cases, we take the total duration as the budget for all the considered test case prioritization methods to ensure a fair comparison.

\begin{table}[t!]
    \centering
    \caption{The linear correlation analysis between $WER$ and $\mathit{CER}$ across different models and datasets.}
    \begin{tabular}{ccc}
    \hline
    Models     & Coefficient & $p$-value \\ \hline
    QuartzNet   & 0.996       & $<0.01$   \\
    HuBERT      & 0.997       & $<0.01$   \\
    wav2vec 2.0 & 0.996       & $<0.01$   \\ \hline
    \end{tabular}
    \label{tab:RQ1-correlation}
    \end{table}

The ASR performance on the selected test suites, i.e., $WER$ and $\mathit{CER}$, can quantify the effectiveness of a test case prioritization method.
We conduct a linear correlation analysis between $WER$ and $\mathit{CER}$ and present the results in Table~\ref{tab:RQ1-correlation}.
It shows that there is a strong linear correlation between the $WER$ and $\mathit{CER}$ across different models and datasets. 
In the following parts of the paper, therefore, we just use $WER$ to measure the performance of ASR systems. 
We compute the relative improvement over a simple baseline: random selection. Assuming that the ASR performance on the randomly chosen test suites and the test suites chosen from the method $A$ are $WER_{rnd}$ and $WER_A$, we use $\frac{WER_A - WER_{rnd}}{WER_{rnd}}$ to measure the relative improvement achieved by method $A$ over the random method.

To further interpret the results, we visualize them by plotting each method's {\em average improvement} in Figure~\ref{fig:RQ1}. 
For each pair of methods and datasets, we ran the method on the dataset 12 times: 3 random seeds and 4 budget settings. We compute a relative improvement value for each of the 12 selected suites over the random selection. We use the average of the 12 relative improvement values as the {\em average improvement}. 
According to the types of ASR systems, we divide the average improvement results into three groups, presented in Figure~\ref{fig:RQ1} (a), (b), and (c), respectively. 
\toolname achieves greater improvement than others, which suggests that it is more effective.
We find that our method can achieve the best performance across all the three models and almost all the datasets (35 out of 36). 
On average, the $WER$ on test suites prioritized by \toolname is $12.63\%$ higher than that prioritized by the second-best method, \emph{CrossASR}.
We conduct the Wilcoxon signed-rank tests, showing that our method is statistically significantly better than all other methods ($p$-value less than 0.05). 
Besides, the method~\cite{6947957} that enhances phoneme richness significantly performs worse than other methods.

\vspace{0.2cm}
\ans{\textbf{Answers to RQ1:} \toolname is the most effective method to prioritize test cases. On average, \toolname can uncover $12.63\%$ more word errors than the best-performing baseline under the same testing budget.}

\subsection*{RQ2. How is the performance of \toolname in enhancing ASR systems performance?}
\label{subsec:RQ2}
This research question explores the value of using selected suites to improve the performance of ASR systems. 
After fine-tuning an ASR system on the test suites obtained from RQ1, we evaluate the original models and fine-tuned models on the same test set to analyze the impact of different selected suites on the model performance changes. 

We compute the relative difference of the $WER$ after the model is fine-tuned on the selected suite. 
We denote the $WER$ of the original model as $WER_{ori}$ and the $WER$ of the model fine-tuned on test suite selected by method A as $WER_{A}$. 
We use $\frac{WER_{orig} - WER_{A}}{WER_{orig}}$ to measure the relative improvement. 
A more effective method should lead to a greater relative improvement.
Table~\ref{tab:RQ2} shows the average improvements of the models fine-tuned on the test suites obtained from different methods. We can observe that unlike the results in RQ1, none of the methods can always outperform others across different datasets and models.
In Table~\ref{tab:RQ2}, we use \textcolor{flatgreen}{$\blacksquare$} blocks to label the methods leading to the largest improvement and \textcolor{flatred}{$\blacksquare$} blocks to label the methods leading to the smallest improvement.
Overall, \toolname achieves the largest improvement for 11 cases (out of 36), followed by \emph{CrossASR} (7 cases). 
Furthermore, we run 30 statistical tests to validate hypotheses of the following form:

\begin{quote}
    \textit{The ASR performance improvement by fine-tuning models using data prioritized by $method_A$ is significantly larger than that of $method_B$.}
\end{quote}

The $method_A$ and $method_B$ are any two pairs of investigated methods. 
To test the hypothesis, we conduct a Wilcoxon signed-rank test for each pair of methods to investigate whether the improvement given by one method is significantly better than the improvement brought by the other method. 
The statistical test results are displayed in the last row of Table~\ref{tab:RQ2}. 
A difference is viewed as significant if the $p$-value is less than $0.05$.
We observe that our proposed \toolname is significantly better than 4 other methods, followed by \emph{CrossASR} which significantly outperforms 3 methods. 
In all the investigated methods, none of them is significantly better than \toolnamenospace.

\begin{table}[!t]
    \caption{The average improvement of different methods against the original ASR system performance. The \textcolor{flatred}{$\blacksquare$} blocks mean that a method achieves the smallest improvement, and the \textcolor{flatgreen}{$\blacksquare$} blocks mean that a method achieves the largest improvement. `PR', `CASR', and `Ours' stand for Phoneme-Rich~\cite{phonme_rich}, \emph{CrossASR}~\cite{crossasr}, and \toolnamenospace, respectively. The last row shows the statistical test results. For example, `e' means that a method is statistically better than \emph{CrossASR}.}
    \begin{tabular}{ccccccc}
    \hline
    \textbf{Dataset} & \textbf{Random} & \textbf{PR} & \textbf{\emph{PEP}} & \textbf{\emph{PEP-D}}   & \textbf{ \emph{CASR}}   & \textbf{Ours}    \\ 
        \hline
        YBAA &                 20.60 &   \setBadColor{}19.37 &                 20.65 &                 20.38 &                 20.15 &  \setGoodColor{}20.77 \\
        ZHAA &                 17.51 &                 17.54 &                 18.58 &   \setBadColor{}13.93 &  \setGoodColor{}19.11 &                 16.42 \\
        ASI  &                 14.47 &   \setBadColor{}10.24 &                 15.96 &  \setGoodColor{}16.01 &                 15.64 &                 15.85 \\
        TNI  &  \setGoodColor{}13.13 &                 12.10 &    \setBadColor{}9.93 &                 11.80 &                 11.78 &                 12.98 \\
        NCC  &                 21.05 &                 21.38 &  \setGoodColor{}21.94 &                 21.62 &   \setBadColor{}20.91 &                 21.09 \\
        TXHC &                 28.77 &   \setBadColor{}26.91 &                 28.65 &                 28.43 &                 28.06 &  \setGoodColor{}29.55 \\
        EBVS &                 11.58 &                 10.86 &   \setBadColor{}10.57 &                 11.53 &  \setGoodColor{}12.20 &                 11.50 \\
        ERMS &                 18.96 &                 19.91 &                 19.96 &  \setGoodColor{}20.25 &   \setBadColor{}18.29 &                 19.11 \\
        YDCK &                 16.09 &   \setBadColor{}13.73 &                 17.09 &                 15.63 &                 17.30 &  \setGoodColor{}19.36 \\
        YKWK &                 20.78 &   \setBadColor{}18.39 &                 21.63 &  \setGoodColor{}22.46 &                 19.41 &                 20.91 \\
        THV  &   \setBadColor{}18.16 &                 18.30 &                 19.77 &                 18.97 &                 19.92 &  \setGoodColor{}20.19 \\
        TLV  &                 24.49 &  \setGoodColor{}25.84 &                 23.94 &   \setBadColor{}23.40 &                 24.84 &                 24.16 \\
        \hline
        YBAA &                 22.15 &   22.53 &                 22.39 &                 \setBadColor{}16.91 &                 21.14 &  \setGoodColor{}23.00 \\
        ZHAA &                 22.35 &   \setBadColor{}19.89 &                 23.49 &                 23.37 &                 22.52 &  \setGoodColor{}23.75 \\
        ASI  &   \setBadColor{}13.19 &                 14.84 &                 14.30 &                 13.79 &                 14.21 &  \setGoodColor{}15.78 \\
        TNI  &                 22.23 &                 21.61 &   \setBadColor{}17.65 &  \setGoodColor{}23.21 &                 23.03 &                 22.44 \\
        NCC  &  \setGoodColor{}24.55 &                 24.23 &                 22.68 &                 21.28 &   \setBadColor{}21.05 &                 23.58 \\
        TXHC &                 21.00 &   \setBadColor{}19.07 &  \setGoodColor{}23.92 &                 21.02 &                 23.70 &                 21.36 \\
        EBVS &                 20.81 &  \setGoodColor{}22.22 &                 21.93 &   \setBadColor{}18.01 &                 21.72 &                 22.15 \\
        ERMS &                 27.79 &                 27.60 &                 28.27 &   \setBadColor{}21.84 &                 28.25 &  \setGoodColor{}29.27 \\
        YDCK &                 22.53 &   \setBadColor{}21.65 &                 22.46 &  \setGoodColor{}25.66 &                 23.46 &                 23.39 \\
        YKWK &   \setBadColor{}30.75 &                 32.40 &                 31.39 &                 31.24 &  \setGoodColor{}35.00 &                 33.53 \\
        THV  &                 26.18 &                 26.47 &  \setGoodColor{}26.50 &   \setBadColor{}24.32 &                 26.44 &                 26.42 \\
        TLV  &                 32.95 &                 33.10 &                 32.72 &   \setBadColor{}29.21 &                 31.79 &  \setGoodColor{}33.59 \\
        \hline
        YBAA &                 13.03 &   \setBadColor{}8.50 &                 10.72 &  \setGoodColor{}13.08 &                 11.47 &                 11.05 \\
        ZHAA &                 14.51 &  \setBadColor{}13.73 &  \setGoodColor{}15.64 &                 14.69 &                 14.90 &                 13.89 \\
        ASI  &                 18.52 &                17.36 &   \setBadColor{}16.62 &                 17.63 &                 18.95 &  \setGoodColor{}19.84 \\
        TNI  &                 30.04 &                29.65 &                 29.61 &   \setBadColor{}29.24 &                 31.41 &  \setGoodColor{}31.67 \\
        NCC  &  \setGoodColor{}22.20 &                22.14 &                 20.81 &                 20.92 &   \setBadColor{}20.48 &                 21.14 \\
        TXHC &  24.67 &  \setBadColor{}21.55 &                 22.13 &                 23.48 &                 \setGoodColor{}24.68 &                 22.37 \\
        EBVS &  \setGoodColor{}15.05 &  \setBadColor{}14.00 &                 14.14 &                 14.39 &                 14.42 &                 14.54 \\
        ERMS &                 26.83 &                27.06 &   \setBadColor{}26.60 &                 26.74 &  \setGoodColor{}28.78 &                 28.02 \\
        YDCK &                 25.31 &                24.19 &                 25.72 &   \setBadColor{}24.09 &  \setGoodColor{}26.89 &                 26.34 \\
        YKWK &                 13.07 &                13.15 &                 11.59 &   \setBadColor{}11.07 &  \setGoodColor{}13.49 &                 13.04 \\
        THV  &                 19.65 &                18.85 &  \setGoodColor{}20.55 &                 19.09 &                 19.60 &   \setBadColor{}18.08 \\
        TLV  &  \setGoodColor{}22.98 &                21.22 &                 21.76 &   \setBadColor{}21.19 &                 21.57 &                 22.09 \\
        \hline
        \textbf{Results}      & $bd$   & -      &  $bd$   &  - & $bcd$  &  $abcd$ \\
        \hline

    \end{tabular}
    \label{tab:RQ2}
    \caption*{
    a: $>$ Random, b: $>$ Phoneme-rich,  c: $>$ \emph{PEP},  d: $>$ \emph{PEP-D}, \\  e: $>$  \emph{CrossASR},  f: $>$ \toolnamenospace
    }
    \vspace{-0.7cm}
\end{table}

\vspace{0.2cm}
\ans{
    \textbf{Answers to RQ2}: In terms of prioritizing test cases to improve model performance, \toolname performs significantly better than 4 evaluated methods, followed by \emph{CrossASR} which significantly outperforms 3 methods.
}


%

\subsection*{RQ3. What kind of datasets are more effective in improving ASR system performance?}
This research question bridges the results of RQ1 and RQ2. 
In RQ2, we fine-tune ASR systems on the test suites selected in RQ1, which are called the \emph{fine-tuning set}.
We explore what features of the fine-tuning sets affect the improvement of ASR systems the most. 
We consider the following three features: 
\begin{enumerate}
    \item the original model's $WER$ on the fine-tuning set. This feature corresponds to the objective of \toolnamenospace, i.e., selecting test cases that uncover more word errors. 
    \item Euclidean distance between the phoneme distribution of the fine-tuning set and the 
    uniform distribution. This feature measures the triphone richness~\cite{6947957} of a fine-tuning set. 
    \item the value of the submodular function in Equation~\ref{eq:utility}. This feature measures phoneme diversity of a fine-tuning set.
\end{enumerate}
We use Spearman correlation analysis~\cite{kotz2005encyclopedia} to identify the correlation between features and performance improvement.

We first analyze the correlation between the feature (1), i.e., original model's $WER$ on the fine-tuning set, and the ASR performance improvement. 
To control the data size at a similar level, we divide the data by the dataset, ASR system, and budget to get 144 groups; each group contains 18 data points (i.e., 6 methods and 3 seeds). 
In each group, we conduct the Spearman correlation analysis between the feature and the relative improvement.
We find significant correlations in 35 groups ($p$-value $< 0.05$). 
According to the definition of correlation in Guildford scale~\cite{guilford1950fundamental}, an absolute value smaller than 0.4 indicates a weak correlation; an absolute value between 0.4 and 0.7 means a moderate correlation; and otherwise, the correlation is high (0.7-0.9) or very high (above 0.9).
We find high correlations in 55\% (19/35) groups, moderate correlations in 34\% (12/35) groups and weak correlations in 11\% (4/35) groups. 
In the 144 groups, 99 show positive correlations, and the remaining 45 show negative correlations. 
Of the 99 positive correlations, 27 of them are significant, while only 2 of the 45 negative correlations are significant. Of all the 29 significant correlations, 28 are moderate, and 1 is high. 

We then analyze the distance between triphone distribution and the uniform distribution. 
We find that only 12.5\% of groups show significant correlations; 13 are positive and 5 are negative. The number of groups with significant correlations between the submodular function and the performance improvement is even fewer: only 7.64\% (i.e., 11 groups) from the 144 groups are significant. Eight correlations are negative and three correlations are positive. 

The above results show that the ASR system's $WER$ on the fine-tuning set is more important than the two metrics (measuring the phoneme and triphone richness) in impacting the model performance. 
It also explains why our method \toolnamenospace, which can better prioritize test cases with stronger ability to uncover errors, is also significantly better than the method that aims to increase the phoneme richness.

\vspace*{0.2cm}
\ans{
    \textbf{Answers to RQ3}: The original model's $WER$ on the training data shows a stronger (positive) correlation than the other two features about diversity, suggesting that $WER$ is a better indicator of a fine-tuning set's potential to improve ASR system performance.
}

\section{Discussion}
\label{sec:discussion}
This section discusses the impact of considering phoneme and triphone diversity, as well as the threats to validity.

\subsection{The Impact of Considering Diversity}

Awasthi et al.~\cite{icassp2021} measuring the importance of a test suite from two aspects, (1) the ability to uncover errors, (2) phoneme diversity. 
The intuition behind is twofold.
First, diverse test cases can uncover diverse word errors (i.e., the reasons for a test case to fail). 
Similar to other learning-based models, the training and repair of ASR systems also follow a data-driven paradigm. 
Using more diverse data to fine-tune an ASR system may repair more miscellaneous errors and thus better improve model performance. 
Assuming that the word `{\tt love}' will be wrongly recognized, one can build a very large collection of test cases containing words like `{\tt lovely}', `{\tt loved}', and so on. However, if we use these test cases to retrain a model, we may only be able to repair limited word errors, i.e., word errors associated with `{\tt love}'. If we have a more diverse test suite of the same size, word errors beyond `{\tt love}' may get repaired via retraining, leading to more significant performance improvement.

Secondly, the performance improvement of ASR systems often shows diminishing returns with increasing amounts of training data~\cite{lin2009select,7178848,6854213}. For a specific word error, if we have already collected some relevant data to repair it, additionally collecting more data related to the same error cannot bring proportional performance improvement. 
However, the results from RQ3 suggest that diversity is not the main feature that can contribute to larger model performance improvement. 
Thus, we further analyze the impact of diversity in ASR testing.

Awasthi et al.~\cite{icassp2021} combine the diversity-enhancing function with the \emph{PEP} and shows that its performance can outperform a method that only considers test case diversity. 
However, it is still unclear how the additional diversity-enhancing function affects the performance of PEP. 
Thus, we conduct an ablation study to compare \emph{PEP} and \emph{PEP-D}.
In Figure~\ref{fig:RQ1}, we can observe that considering diversity harms the effectiveness of \emph{PEP}, especially on the two Transformer-based models HuBERT and wav2vec 2.0. 
In the last row in Table~\ref{tab:RQ2}, we can find that the performance improvement brought by \emph{PEP} is significantly larger than that brought by its diversity-enhancing variant.
The results also suggest that considering diversity can cause negative impacts on both prioritizing test cases to uncover ASR errors and improving ASR systems. 
It is also worthy of validating the findings on larger datasets and other models as future work. 

\subsection{Threats to Validity}
\noindent \textbf{Threats to internal validity.} As stated in Section~\ref{subsec:models_datasets}, for ASR systems we use their official implementation and the corresponding pre-trained models released by the authors. We use the repositories provided by the authors to implement the error predictor in \emph{CrossASR}. However, Shinohara and Awasthi et al.'s works~\cite{icassp2021,phonme_rich} are not accompanied by any corresponding open-source implementation. We implement the two methods by ourselves and test the code to mitigate the threats to internal validity. 

\vspace{0.1cm} 
\noindent \textbf{Threats to external validity.} To minimize the threats, we evaluate 6 selection methods on 3 ASR systems and 12 datasets of different accents, using 4 budget settings and 3 random seeds. 
Still, the results may not be generalizable to other models and datasets. We encourage researchers to extend this study further with more models and datasets.

\vspace{0.1cm} 
\noindent \textbf{Threats to construct validity.} Even for the same ASR system, its performance on different datasets can be greatly different. When we compare the ability to improve model performance (RQ2), thus we compute the average value of relative improvement rather than the absolute changes in ASR performance.

\section{Related Work}
\label{sec:related_work}

This section presents the previous studies about testing ASR systems (Section~\ref{subsec:ASR-test}) and other DNN-based systems (Section~\ref{subsec:DNN-test}).

\subsection{ASR System Testing}
\label{subsec:ASR-test}

As pointed out by Zhang et al.~\cite{ml-testing-survey}, there remain open exciting testing research opportunities in speech recognition. Du et al.~\cite{DeepCruiser} proposed DeepCruiser that defines a set of test coverage criteria specialized for RNN-based models. 
Although it is evaluated on a RNN-based ASR system, DeepCruiser is inapplicable to the state-of-the-art ASR systems that are based Transformer~\cite{wav2vec2,HuBERT}. 
Besides, DeepCruiser requires the parameter information to compute the test coverage criteria, but this paper assumes a practical black-box testing scenario. 
Asyroﬁ et al.~\cite{crossasr} propose \emph{CrossASR}, a differential testing framework for ASR systems. \emph{CrossASR} can also be applied in black-box settings and utilizes text-to-speech systems to synthesize test cases. The nature of differential testing requires the framework to access other high-quality ASR systems as references. To facilitate the testing process, it leverages a failure estimator. \emph{CrossASR}++~\cite{crossasrpp} is then proposed as an extension of \emph{CrossASR}. Asyroﬁ et al.~\cite{asrevolve} use the generated test cases to retrain ASR systems and find that they can improve the model performance on synthetic datasets.

Wu and Rajan~\cite{catchme} generate speech test cases by manipulating the audio signal using a psychoacoustic model that maintains the audio perturbations below the thresholds of human perception. 
Ji et al.~\cite{ASRTest} adopt Gini impurity~\cite{DeepGini} to generate speech test cases and improve testing efficiency.
Rajan et al.~\cite{aequevox} focus on the fairness of ASR systems. They propose a method to identify words that are not robust to simulated environment noises and find that ASR systems are biased towards non-native and female speakers.

This paper evaluates the works on prioritizing ASR test cases based on texts. Here we briefly introduce the works on prioritizing test cases based on audio, aiming at finding the most informative audio from a large corpus of speech data. One branch of works in this direction is to use the uncertainty-based sampling method. The intuition is to select the input that a model is most uncertain about. The uncertainty can be measured using confidence score~\cite{4960685}, entropy~\cite{confidence}, etc. The active learning methods usually assume access to an existing model and a small amount of labeled data.
For example, Malhotra et al.~\cite{malhotra2019active} leverage active learning to select a representative dataset from low-resources domains. The Submodular function can also be defined on the audio-related information. Lin and Bilmes~\cite{LinB09-3} combine a submodular function and active learning to select a subset of training data for ASR systems. 
These works assume the existence of audio, which might be unavailable in some cases, e.g., the testing scenario 1 presented in Section~\ref{sec:intro}. It will be worth exploring to select test cases based on audio in the future.

\subsection{DNN Test Case Generation and Selection}
\label{subsec:DNN-test}
With the wide adoption of DNN models in different domains, the testing and quality assurance of DNN models recently have attracted much attention from both software engineering and artificial intelligence communities. 
Researchers propose to test different aspects of DNN models, including correctness~\cite{crossasr,crossasrpp}, robustness~\cite{DeepHunter}, fairness~\cite{biasfinder}, etc.
Different methods are designed to test DNN models for various tasks, e.g., vulnerable code detection~\cite{10.1145/3510003.3510146}, reinforcement learning~\cite{10.1145/3564625.3564636}, etc.
We refer interested readers to the paper by Zhang et al.~\cite{ml-testing-survey} for a comprehensive survey of machine learning testing works. Here we discuss the test case generation and selection methods that are most relevant to this study.

As an analogy to the code coverage-driven testing in conventional software systems, Pei et al.~\cite{DeepXplore} propose DeepXplore, a tool that generates DNN test cases by optimizing neuron coverage, to uncover the wrong behaviors in DNN. 
A sequence of works follows this direction and proposes structural neuron coverage metrics~\cite{DeepHunter,DeepTest,DeepGauge,DeepCT} to generate DNN test cases. Recently, researchers~\cite{yang2022revisiting,harel2020neuron,FSE_Yan,ICECCS} have demonstrated that the existing coverage-driven methods are not effective in producing high-quality test cases. Hu et al.~\cite{hu_tosem} and Ma et al.~\cite{ma_tosem} conducted empirical studies on different test case selection methods on computer vision and natural language systems. 
As mentioned in Section~\ref{subsec:baseline}, the existing metrics are not directly applicable to ASR testing scenarios in this paper, which aims to select test cases from texts in a black-box manner. 

\section{Conclusion and Future Work}
\label{sec:conclusion}

This paper proposes \toolnamenospace, a tool that can predict which words in an ASR test case are likely to be wrongly recognized only based on the reference text. 
Based on \toolnamenospace, we propose a method to prioritize speech test cases.
We conduct large-scale experiments and obtained \numbermodel ASR systems to compare the proposed method with other baselines. 
On average, the $\mathit{WER}$ on test suites prioritized by \toolname is $12.63\%$ higher than that prioritized by the second-best method.
We fine-tune ASR systems on the test cases obtained from different methods to investigate the value of using selected suites to enhance the model performance. 
The statistical testing shows that the model improvement brought by our method is significantly better than four evaluated methods, and none of the investigated methods is significantly better than our proposed method.
We conduct additional statistical testing to analyze what features of a fine-tuning set are correlated with the model improvement. 
The original model's $\mathit{WER}$ on the fine-tuning set shows a stronger correlation than the other two features about the triphone and phoneme richness.
It suggests that $WER$ is a better indicator of a fine-tuning set's potential to improve ASR system performance.
It also explains why our proposed \toolnamenospace, which can select more erroneous test cases, is also significantly better than the method that aims at increasing the phoneme richness.

In the future, we plan to extend the study to even more ASR systems and larger datasets. 
We also plan to explore the value of using synthetic audio to improve ASR systems.

\vspace{0.2cm}
\noindent \textbf{Replication Package}: We release the replication package online to facilitate future research.\footnote{\url{https://github.com/yangzhou6666/ASRProphet}}

\ifCLASSOPTIONcompsoc
  \section*{Acknowledgments}
\else
  \section*{Acknowledgment}
\fi

This research is supported by the Ministry of Education, Singapore under its Academic Research Fund Tier 3 (Award ID: MOET32020-0004). Any opinions, findings and conclusions or recommendations expressed in this material are those of the author(s) and do not reflect the views of the Ministry of Education, Singapore.

\balance 
\bibliographystyle{IEEEtran}
\bibliography{reference}


\begin{thebibliography}{66}


\ifx \showCODEN    \undefined \def \showCODEN     #1{\unskip}     \fi
\ifx \showDOI      \undefined \def \showDOI       #1{#1}\fi
\ifx \showISBNx    \undefined \def \showISBNx     #1{\unskip}     \fi
\ifx \showISBNxiii \undefined \def \showISBNxiii  #1{\unskip}     \fi
\ifx \showISSN     \undefined \def \showISSN      #1{\unskip}     \fi
\ifx \showLCCN     \undefined \def \showLCCN      #1{\unskip}     \fi
\ifx \shownote     \undefined \def \shownote      #1{#1}          \fi
\ifx \showarticletitle \undefined \def \showarticletitle #1{#1}   \fi
\ifx \showURL      \undefined \def \showURL       {\relax}        \fi
\providecommand\bibfield[2]{#2}
\providecommand\bibinfo[2]{#2}
\providecommand\natexlab[1]{#1}
\providecommand\showeprint[2][]{arXiv:#2}

\bibitem[\protect\citeauthoryear{Ardila, Branson, Davis, Kohler, Meyer,
  Henretty, Morais, Saunders, Tyers, and Weber}{Ardila et~al\mbox{.}}{2020}]%
        {common-voice}
\bibfield{author}{\bibinfo{person}{Rosana Ardila}, \bibinfo{person}{Megan
  Branson}, \bibinfo{person}{Kelly Davis}, \bibinfo{person}{Michael Kohler},
  \bibinfo{person}{Josh Meyer}, \bibinfo{person}{Michael Henretty},
  \bibinfo{person}{Reuben Morais}, \bibinfo{person}{Lindsay Saunders},
  \bibinfo{person}{Francis Tyers}, {and} \bibinfo{person}{Gregor Weber}.}
  \bibinfo{year}{2020}\natexlab{}.
\newblock \showarticletitle{Common Voice: A Massively-Multilingual Speech
  Corpus}. In \bibinfo{booktitle}{\emph{Proceedings of the 12th Language
  Resources and Evaluation Conference}}. \bibinfo{publisher}{European Language
  Resources Association}, \bibinfo{address}{Marseille, France},
  \bibinfo{pages}{4218--4222}.
\newblock
\showISBNx{979-10-95546-34-4}
\urldef\tempurl%
\url{https://aclanthology.org/2020.lrec-1.520}
\showURL{%
\tempurl}


\bibitem[\protect\citeauthoryear{Asyrofi, Thung, Lo, and Jiang}{Asyrofi
  et~al\mbox{.}}{2020}]%
        {crossasr}
\bibfield{author}{\bibinfo{person}{Muhammad~Hilmi Asyrofi},
  \bibinfo{person}{Ferdian Thung}, \bibinfo{person}{David Lo}, {and}
  \bibinfo{person}{Lingxiao Jiang}.} \bibinfo{year}{2020}\natexlab{}.
\newblock \showarticletitle{CrossASR: Efficient Differential Testing of
  Automatic Speech Recognition via Text-To-Speech}. In
  \bibinfo{booktitle}{\emph{2020 IEEE International Conference on Software
  Maintenance and Evolution (ICSME)}}. \bibinfo{pages}{640--650}.
\newblock
\urldef\tempurl%
\url{https://doi.org/10.1109/ICSME46990.2020.00066}
\showDOI{\tempurl}


\bibitem[\protect\citeauthoryear{Asyrofi, Yang, and Lo}{Asyrofi
  et~al\mbox{.}}{2021a}]%
        {crossasrpp}
\bibfield{author}{\bibinfo{person}{Muhammad~Hilmi Asyrofi},
  \bibinfo{person}{Zhou Yang}, {and} \bibinfo{person}{David Lo}.}
  \bibinfo{year}{2021}\natexlab{a}.
\newblock \showarticletitle{CrossASR++: A Modular Differential Testing
  Framework for Automatic Speech Recognition}. In
  \bibinfo{booktitle}{\emph{Proceedings of the 29th ACM Joint Meeting on
  European Software Engineering Conference and Symposium on the Foundations of
  Software Engineering}} (Athens, Greece) \emph{(\bibinfo{series}{ESEC/FSE
  2021})}. \bibinfo{publisher}{Association for Computing Machinery},
  \bibinfo{address}{New York, NY, USA}, \bibinfo{pages}{1575–1579}.
\newblock
\showISBNx{9781450385626}
\urldef\tempurl%
\url{https://doi.org/10.1145/3468264.3473124}
\showDOI{\tempurl}


\bibitem[\protect\citeauthoryear{Asyrofi, Yang, Shi, Quan, and Lo}{Asyrofi
  et~al\mbox{.}}{2021b}]%
        {asrevolve}
\bibfield{author}{\bibinfo{person}{Muhammad~Hilmi Asyrofi},
  \bibinfo{person}{Zhou Yang}, \bibinfo{person}{Jicke Shi},
  \bibinfo{person}{Chu~Wei Quan}, {and} \bibinfo{person}{David Lo}.}
  \bibinfo{year}{2021}\natexlab{b}.
\newblock \showarticletitle{Can Differential Testing Improve Automatic Speech
  Recognition Systems?}. In \bibinfo{booktitle}{\emph{2021 IEEE International
  Conference on Software Maintenance and Evolution (ICSME)}}.
  \bibinfo{pages}{674--678}.
\newblock
\urldef\tempurl%
\url{https://doi.org/10.1109/ICSME52107.2021.00079}
\showDOI{\tempurl}


\bibitem[\protect\citeauthoryear{Asyrofi, Yang, Yusuf, Kang, Thung, and
  Lo}{Asyrofi et~al\mbox{.}}{2021c}]%
        {biasfinder}
\bibfield{author}{\bibinfo{person}{Muhammad~Hilmi Asyrofi},
  \bibinfo{person}{Zhou Yang}, \bibinfo{person}{Imam Nur~Bani Yusuf},
  \bibinfo{person}{Hong~Jin Kang}, \bibinfo{person}{Ferdian Thung}, {and}
  \bibinfo{person}{David Lo}.} \bibinfo{year}{2021}\natexlab{c}.
\newblock \showarticletitle{BiasFinder: Metamorphic Test Generation to Uncover
  Bias for Sentiment Analysis Systems}.
\newblock \bibinfo{journal}{\emph{IEEE Transactions on Software Engineering}}
  (\bibinfo{year}{2021}), \bibinfo{pages}{1--1}.
\newblock
\urldef\tempurl%
\url{https://doi.org/10.1109/TSE.2021.3136169}
\showDOI{\tempurl}


\bibitem[\protect\citeauthoryear{Awasthi, Kansal, Sarawagi, and Jyothi}{Awasthi
  et~al\mbox{.}}{2021}]%
        {icassp2021}
\bibfield{author}{\bibinfo{person}{Abhijeet Awasthi}, \bibinfo{person}{Aman
  Kansal}, \bibinfo{person}{Sunita Sarawagi}, {and} \bibinfo{person}{Preethi
  Jyothi}.} \bibinfo{year}{2021}\natexlab{}.
\newblock \showarticletitle{Error-Driven Fixed-Budget ASR Personalization for
  Accented Speakers}. In \bibinfo{booktitle}{\emph{ICASSP 2021 - 2021 IEEE
  International Conference on Acoustics, Speech and Signal Processing
  (ICASSP)}}. \bibinfo{pages}{7033--7037}.
\newblock
\urldef\tempurl%
\url{https://doi.org/10.1109/ICASSP39728.2021.9414830}
\showDOI{\tempurl}


\bibitem[\protect\citeauthoryear{Azeemi, Qazi, and Raza}{Azeemi
  et~al\mbox{.}}{2022}]%
        {DBLP:journals/corr/abs-2203-09829}
\bibfield{author}{\bibinfo{person}{Abdul~Hameed Azeemi},
  \bibinfo{person}{Ihsan~Ayyub Qazi}, {and} \bibinfo{person}{Agha~Ali Raza}.}
  \bibinfo{year}{2022}\natexlab{}.
\newblock \showarticletitle{Towards Representative Subset Selection for
  Self-Supervised Speech Recognition}.
\newblock \bibinfo{journal}{\emph{CoRR}}  \bibinfo{volume}{abs/2203.09829}
  (\bibinfo{year}{2022}).
\newblock
\urldef\tempurl%
\url{https://doi.org/10.48550/arXiv.2203.09829}
\showDOI{\tempurl}
\showeprint[arXiv]{2203.09829}


\bibitem[\protect\citeauthoryear{Baevski, Zhou, Mohamed, and Auli}{Baevski
  et~al\mbox{.}}{2020}]%
        {wav2vec2}
\bibfield{author}{\bibinfo{person}{Alexei Baevski}, \bibinfo{person}{Yuhao
  Zhou}, \bibinfo{person}{Abdelrahman Mohamed}, {and} \bibinfo{person}{Michael
  Auli}.} \bibinfo{year}{2020}\natexlab{}.
\newblock \showarticletitle{wav2vec 2.0: A framework for self-supervised
  learning of speech representations}.
\newblock \bibinfo{journal}{\emph{Advances in Neural Information Processing
  Systems}}  \bibinfo{volume}{33} (\bibinfo{year}{2020}),
  \bibinfo{pages}{12449--12460}.
\newblock


\bibitem[\protect\citeauthoryear{Chen, Lou, Zhang, Zhou, Wang, Hao, and
  Zhang}{Chen et~al\mbox{.}}{2018}]%
        {10.1145/3236024.3236053}
\bibfield{author}{\bibinfo{person}{Junjie Chen}, \bibinfo{person}{Yiling Lou},
  \bibinfo{person}{Lingming Zhang}, \bibinfo{person}{Jianyi Zhou},
  \bibinfo{person}{Xiaoleng Wang}, \bibinfo{person}{Dan Hao}, {and}
  \bibinfo{person}{Lu Zhang}.} \bibinfo{year}{2018}\natexlab{}.
\newblock \showarticletitle{Optimizing Test Prioritization via Test
  Distribution Analysis}. In \bibinfo{booktitle}{\emph{Proceedings of the 2018
  26th ACM Joint Meeting on European Software Engineering Conference and
  Symposium on the Foundations of Software Engineering}} (Lake Buena Vista, FL,
  USA) \emph{(\bibinfo{series}{ESEC/FSE 2018})}.
  \bibinfo{publisher}{Association for Computing Machinery},
  \bibinfo{address}{New York, NY, USA}, \bibinfo{pages}{656–667}.
\newblock
\showISBNx{9781450355735}
\urldef\tempurl%
\url{https://doi.org/10.1145/3236024.3236053}
\showDOI{\tempurl}


\bibitem[\protect\citeauthoryear{Cheng, Zhang, Marinov, and Xu}{Cheng
  et~al\mbox{.}}{2021}]%
        {10.1145/3460319.3464810}
\bibfield{author}{\bibinfo{person}{Runxiang Cheng}, \bibinfo{person}{Lingming
  Zhang}, \bibinfo{person}{Darko Marinov}, {and} \bibinfo{person}{Tianyin Xu}.}
  \bibinfo{year}{2021}\natexlab{}.
\newblock \showarticletitle{Test-Case Prioritization for Configuration
  Testing}. In \bibinfo{booktitle}{\emph{Proceedings of the 30th ACM SIGSOFT
  International Symposium on Software Testing and Analysis}} (Virtual, Denmark)
  \emph{(\bibinfo{series}{ISSTA 2021})}. \bibinfo{publisher}{Association for
  Computing Machinery}, \bibinfo{address}{New York, NY, USA},
  \bibinfo{pages}{452–465}.
\newblock
\showISBNx{9781450384599}
\urldef\tempurl%
\url{https://doi.org/10.1145/3460319.3464810}
\showDOI{\tempurl}


\bibitem[\protect\citeauthoryear{Cooper, Ingebretsen, and Strayer}{Cooper
  et~al\mbox{.}}{2014}]%
        {cooper2014mental}
\bibfield{author}{\bibinfo{person}{Joel~M Cooper}, \bibinfo{person}{Hailey
  Ingebretsen}, {and} \bibinfo{person}{David~L Strayer}.}
  \bibinfo{year}{2014}\natexlab{}.
\newblock \showarticletitle{Mental workload of common voice-based vehicle
  interactions across six different vehicle systems}.
\newblock  (\bibinfo{year}{2014}).
\newblock


\bibitem[\protect\citeauthoryear{Devlin, Chang, Lee, and Toutanova}{Devlin
  et~al\mbox{.}}{2019}]%
        {bert}
\bibfield{author}{\bibinfo{person}{Jacob Devlin}, \bibinfo{person}{Ming-Wei
  Chang}, \bibinfo{person}{Kenton Lee}, {and} \bibinfo{person}{Kristina
  Toutanova}.} \bibinfo{year}{2019}\natexlab{}.
\newblock \showarticletitle{{BERT}: Pre-training of Deep Bidirectional
  Transformers for Language Understanding}. In
  \bibinfo{booktitle}{\emph{Proceedings of the 2019 Conference of the North
  {A}merican Chapter of the Association for Computational Linguistics: Human
  Language Technologies, Volume 1 (Long and Short Papers)}}.
  \bibinfo{publisher}{Association for Computational Linguistics},
  \bibinfo{address}{Minneapolis, Minnesota}, \bibinfo{pages}{4171--4186}.
\newblock
\urldef\tempurl%
\url{https://doi.org/10.18653/v1/N19-1423}
\showDOI{\tempurl}


\bibitem[\protect\citeauthoryear{Dong, Zhang, Wang, Liu, Sun, Hao, Wang, Wang,
  Dong, and Dai}{Dong et~al\mbox{.}}{2020}]%
        {ICECCS}
\bibfield{author}{\bibinfo{person}{Yizhen Dong}, \bibinfo{person}{Peixin
  Zhang}, \bibinfo{person}{Jingyi Wang}, \bibinfo{person}{Shuang Liu},
  \bibinfo{person}{Jun Sun}, \bibinfo{person}{Jianye Hao},
  \bibinfo{person}{Xinyu Wang}, \bibinfo{person}{Li Wang},
  \bibinfo{person}{Jinsong Dong}, {and} \bibinfo{person}{Ting Dai}.}
  \bibinfo{year}{2020}\natexlab{}.
\newblock \showarticletitle{An Empirical Study on Correlation between Coverage
  and Robustness for Deep Neural Networks}. In \bibinfo{booktitle}{\emph{2020
  25th International Conference on Engineering of Complex Computer Systems
  (ICECCS)}}. \bibinfo{pages}{73--82}.
\newblock
\urldef\tempurl%
\url{https://doi.org/10.1109/ICECCS51672.2020.00016}
\showDOI{\tempurl}


\bibitem[\protect\citeauthoryear{Du, Xie, Li, Ma, Zhao, and Liu}{Du
  et~al\mbox{.}}{2018}]%
        {DeepCruiser}
\bibfield{author}{\bibinfo{person}{Xiaoning Du}, \bibinfo{person}{Xiaofei Xie},
  \bibinfo{person}{Yi Li}, \bibinfo{person}{Lei Ma}, \bibinfo{person}{Jianjun
  Zhao}, {and} \bibinfo{person}{Yang Liu}.} \bibinfo{year}{2018}\natexlab{}.
\newblock \showarticletitle{DeepCruiser: Automated Guided Testing for Stateful
  Deep Learning Systems}.
\newblock \bibinfo{journal}{\emph{CoRR}}  \bibinfo{volume}{abs/1812.05339}
  (\bibinfo{year}{2018}).
\newblock
\showeprint[arXiv]{1812.05339}
\urldef\tempurl%
\url{http://arxiv.org/abs/1812.05339}
\showURL{%
\tempurl}


\bibitem[\protect\citeauthoryear{Elbaum, Malishevsky, and Rothermel}{Elbaum
  et~al\mbox{.}}{2001}]%
        {919106}
\bibfield{author}{\bibinfo{person}{S. Elbaum}, \bibinfo{person}{A.
  Malishevsky}, {and} \bibinfo{person}{G. Rothermel}.}
  \bibinfo{year}{2001}\natexlab{}.
\newblock \showarticletitle{Incorporating varying test costs and fault
  severities into test case prioritization}. In
  \bibinfo{booktitle}{\emph{Proceedings of the 23rd International Conference on
  Software Engineering. ICSE 2001}}. \bibinfo{pages}{329--338}.
\newblock
\urldef\tempurl%
\url{https://doi.org/10.1109/ICSE.2001.919106}
\showDOI{\tempurl}


\bibitem[\protect\citeauthoryear{Feng, Shi, Gao, Wan, Fang, and Chen}{Feng
  et~al\mbox{.}}{2020}]%
        {DeepGini}
\bibfield{author}{\bibinfo{person}{Yang Feng}, \bibinfo{person}{Qingkai Shi},
  \bibinfo{person}{Xinyu Gao}, \bibinfo{person}{Jun Wan},
  \bibinfo{person}{Chunrong Fang}, {and} \bibinfo{person}{Zhenyu Chen}.}
  \bibinfo{year}{2020}\natexlab{}.
\newblock \showarticletitle{DeepGini: Prioritizing Massive Tests to Enhance the
  Robustness of Deep Neural Networks}. In \bibinfo{booktitle}{\emph{Proceedings
  of the 29th ACM SIGSOFT International Symposium on Software Testing and
  Analysis}} (Virtual Event, USA) \emph{(\bibinfo{series}{ISSTA 2020})}.
  \bibinfo{publisher}{Association for Computing Machinery},
  \bibinfo{address}{New York, NY, USA}, \bibinfo{pages}{177–188}.
\newblock
\showISBNx{9781450380089}
\urldef\tempurl%
\url{https://doi.org/10.1145/3395363.3397357}
\showDOI{\tempurl}


\bibitem[\protect\citeauthoryear{Gao, Saha, Prasad, and Roychoudhury}{Gao
  et~al\mbox{.}}{2020}]%
        {sensei}
\bibfield{author}{\bibinfo{person}{Xiang Gao}, \bibinfo{person}{Ripon~K. Saha},
  \bibinfo{person}{Mukul~R. Prasad}, {and} \bibinfo{person}{Abhik
  Roychoudhury}.} \bibinfo{year}{2020}\natexlab{}.
\newblock \showarticletitle{Fuzz Testing Based Data Augmentation to Improve
  Robustness of Deep Neural Networks}. In \bibinfo{booktitle}{\emph{Proceedings
  of the ACM/IEEE 42nd International Conference on Software Engineering}}
  (Seoul, South Korea) \emph{(\bibinfo{series}{ICSE '20})}.
  \bibinfo{publisher}{Association for Computing Machinery},
  \bibinfo{address}{New York, NY, USA}, \bibinfo{pages}{1147–1158}.
\newblock
\showISBNx{9781450371216}
\urldef\tempurl%
\url{https://doi.org/10.1145/3377811.3380415}
\showDOI{\tempurl}


\bibitem[\protect\citeauthoryear{Gong, Yang, Bai, Shi, Sinha, Xu, Lo, Hou, and
  Fan}{Gong et~al\mbox{.}}{2022}]%
        {10.1145/3564625.3564636}
\bibfield{author}{\bibinfo{person}{Chen Gong}, \bibinfo{person}{Zhou Yang},
  \bibinfo{person}{Yunpeng Bai}, \bibinfo{person}{Jieke Shi},
  \bibinfo{person}{Arunesh Sinha}, \bibinfo{person}{Bowen Xu},
  \bibinfo{person}{David Lo}, \bibinfo{person}{Xinwen Hou}, {and}
  \bibinfo{person}{Guoliang Fan}.} \bibinfo{year}{2022}\natexlab{}.
\newblock \showarticletitle{Curiosity-Driven and Victim-Aware Adversarial
  Policies}. In \bibinfo{booktitle}{\emph{Proceedings of the 38th Annual
  Computer Security Applications Conference}} (Austin, TX, USA)
  \emph{(\bibinfo{series}{ACSAC '22})}. \bibinfo{publisher}{Association for
  Computing Machinery}, \bibinfo{address}{New York, NY, USA},
  \bibinfo{pages}{186–200}.
\newblock
\showISBNx{9781450397599}
\urldef\tempurl%
\url{https://doi.org/10.1145/3564625.3564636}
\showDOI{\tempurl}


\bibitem[\protect\citeauthoryear{Guilford}{Guilford}{1950}]%
        {guilford1950fundamental}
\bibfield{author}{\bibinfo{person}{Joy~Paul Guilford}.}
  \bibinfo{year}{1950}\natexlab{}.
\newblock \showarticletitle{Fundamental statistics in psychology and
  education}.
\newblock  (\bibinfo{year}{1950}).
\newblock


\bibitem[\protect\citeauthoryear{Hakkani-T\"{u}r, Riccardi, and
  Tur}{Hakkani-T\"{u}r et~al\mbox{.}}{2006}]%
        {confidence}
\bibfield{author}{\bibinfo{person}{Dilek Hakkani-T\"{u}r},
  \bibinfo{person}{Giuseppe Riccardi}, {and} \bibinfo{person}{Gokhan Tur}.}
  \bibinfo{year}{2006}\natexlab{}.
\newblock \showarticletitle{An Active Approach to Spoken Language Processing}.
\newblock \bibinfo{journal}{\emph{ACM Trans. Speech Lang. Process.}}
  \bibinfo{volume}{3}, \bibinfo{number}{3} (\bibinfo{date}{oct}
  \bibinfo{year}{2006}), \bibinfo{pages}{1–31}.
\newblock
\showISSN{1550-4875}
\urldef\tempurl%
\url{https://doi.org/10.1145/1177055.1177056}
\showDOI{\tempurl}


\bibitem[\protect\citeauthoryear{Harel-Canada, Wang, Gulzar, Gu, and
  Kim}{Harel-Canada et~al\mbox{.}}{2020}]%
        {harel2020neuron}
\bibfield{author}{\bibinfo{person}{Fabrice Harel-Canada},
  \bibinfo{person}{Lingxiao Wang}, \bibinfo{person}{Muhammad~Ali Gulzar},
  \bibinfo{person}{Quanquan Gu}, {and} \bibinfo{person}{Miryung Kim}.}
  \bibinfo{year}{2020}\natexlab{}.
\newblock \showarticletitle{Is Neuron Coverage a Meaningful Measure for Testing
  Deep Neural Networks?}. In \bibinfo{booktitle}{\emph{Proceedings of the 28th
  ACM Joint Meeting on European Software Engineering Conference and Symposium
  on the Foundations of Software Engineering}} (Virtual Event, USA)
  \emph{(\bibinfo{series}{ESEC/FSE 2020})}. \bibinfo{publisher}{Association for
  Computing Machinery}, \bibinfo{address}{New York, NY, USA},
  \bibinfo{pages}{851–862}.
\newblock
\showISBNx{9781450370431}
\urldef\tempurl%
\url{https://doi.org/10.1145/3368089.3409754}
\showDOI{\tempurl}


\bibitem[\protect\citeauthoryear{Hochreiter and Schmidhuber}{Hochreiter and
  Schmidhuber}{1997}]%
        {hochreiter1997long}
\bibfield{author}{\bibinfo{person}{Sepp Hochreiter} {and}
  \bibinfo{person}{J{\"u}rgen Schmidhuber}.} \bibinfo{year}{1997}\natexlab{}.
\newblock \showarticletitle{Long short-term memory}.
\newblock \bibinfo{journal}{\emph{Neural computation}} \bibinfo{volume}{9},
  \bibinfo{number}{8} (\bibinfo{year}{1997}), \bibinfo{pages}{1735--1780}.
\newblock


\bibitem[\protect\citeauthoryear{Hsu, Bolte, Tsai, Lakhotia, Salakhutdinov, and
  Mohamed}{Hsu et~al\mbox{.}}{2021}]%
        {HuBERT}
\bibfield{author}{\bibinfo{person}{Wei-Ning Hsu}, \bibinfo{person}{Benjamin
  Bolte}, \bibinfo{person}{Yao-Hung~Hubert Tsai}, \bibinfo{person}{Kushal
  Lakhotia}, \bibinfo{person}{Ruslan Salakhutdinov}, {and}
  \bibinfo{person}{Abdelrahman Mohamed}.} \bibinfo{year}{2021}\natexlab{}.
\newblock \showarticletitle{HuBERT: Self-Supervised Speech Representation
  Learning by Masked Prediction of Hidden Units}.
\newblock \bibinfo{journal}{\emph{IEEE/ACM Transactions on Audio, Speech, and
  Language Processing}}  \bibinfo{volume}{29} (\bibinfo{year}{2021}),
  \bibinfo{pages}{3451--3460}.
\newblock
\urldef\tempurl%
\url{https://doi.org/10.1109/TASLP.2021.3122291}
\showDOI{\tempurl}


\bibitem[\protect\citeauthoryear{Hu, Guo, Cordy, Xie, Ma, Papadakis, and
  Le~Traon}{Hu et~al\mbox{.}}{2022}]%
        {hu_tosem}
\bibfield{author}{\bibinfo{person}{Qiang Hu}, \bibinfo{person}{Yuejun Guo},
  \bibinfo{person}{Maxime Cordy}, \bibinfo{person}{Xiaofei Xie},
  \bibinfo{person}{Lei Ma}, \bibinfo{person}{Mike Papadakis}, {and}
  \bibinfo{person}{Yves Le~Traon}.} \bibinfo{year}{2022}\natexlab{}.
\newblock \showarticletitle{An Empirical Study on Data Distribution-Aware Test
  Selection for Deep Learning Enhancement}.
\newblock \bibinfo{journal}{\emph{ACM Transactions on Software Engineering and
  Methodology (TOSEM)}} (\bibinfo{year}{2022}).
\newblock


\bibitem[\protect\citeauthoryear{Ji, Feng, Liu, Zhao, and Chen}{Ji
  et~al\mbox{.}}{2022}]%
        {ASRTest}
\bibfield{author}{\bibinfo{person}{Pin Ji}, \bibinfo{person}{Yang Feng},
  \bibinfo{person}{Jia Liu}, \bibinfo{person}{Zhihong Zhao}, {and}
  \bibinfo{person}{Zhenyu Chen}.} \bibinfo{year}{2022}\natexlab{}.
\newblock \showarticletitle{ASRTest: Automated Testing for
  Deep-Neural-Network-Driven Speech Recognition Systems}. In
  \bibinfo{booktitle}{\emph{Proceedings of the 31st ACM SIGSOFT International
  Symposium on Software Testing and Analysis}} (Virtual, South Korea)
  \emph{(\bibinfo{series}{ISSTA 2022})}. \bibinfo{publisher}{Association for
  Computing Machinery}, \bibinfo{address}{New York, NY, USA},
  \bibinfo{pages}{189–201}.
\newblock
\showISBNx{9781450393799}
\urldef\tempurl%
\url{https://doi.org/10.1145/3533767.3534391}
\showDOI{\tempurl}


\bibitem[\protect\citeauthoryear{Kahn, Rivière, Zheng, Kharitonov, Xu,
  Mazaré, Karadayi, Liptchinsky, Collobert, Fuegen, Likhomanenko, Synnaeve,
  Joulin, Mohamed, and Dupoux}{Kahn et~al\mbox{.}}{2020}]%
        {Libri-Light}
\bibfield{author}{\bibinfo{person}{J. Kahn}, \bibinfo{person}{M. Rivière},
  \bibinfo{person}{W. Zheng}, \bibinfo{person}{E. Kharitonov},
  \bibinfo{person}{Q. Xu}, \bibinfo{person}{P.E. Mazaré}, \bibinfo{person}{J.
  Karadayi}, \bibinfo{person}{V. Liptchinsky}, \bibinfo{person}{R. Collobert},
  \bibinfo{person}{C. Fuegen}, \bibinfo{person}{T. Likhomanenko},
  \bibinfo{person}{G. Synnaeve}, \bibinfo{person}{A. Joulin},
  \bibinfo{person}{A. Mohamed}, {and} \bibinfo{person}{E. Dupoux}.}
  \bibinfo{year}{2020}\natexlab{}.
\newblock \showarticletitle{Libri-Light: A Benchmark for ASR with Limited or No
  Supervision}. In \bibinfo{booktitle}{\emph{ICASSP 2020 - 2020 IEEE
  International Conference on Acoustics, Speech and Signal Processing
  (ICASSP)}}. \bibinfo{pages}{7669--7673}.
\newblock
\urldef\tempurl%
\url{https://doi.org/10.1109/ICASSP40776.2020.9052942}
\showDOI{\tempurl}


\bibitem[\protect\citeauthoryear{Kim, Feldt, and Yoo}{Kim
  et~al\mbox{.}}{2019}]%
        {surprise}
\bibfield{author}{\bibinfo{person}{Jinhan Kim}, \bibinfo{person}{Robert Feldt},
  {and} \bibinfo{person}{Shin Yoo}.} \bibinfo{year}{2019}\natexlab{}.
\newblock \showarticletitle{Guiding Deep Learning System Testing Using Surprise
  Adequacy}. In \bibinfo{booktitle}{\emph{Proceedings of the 41st International
  Conference on Software Engineering}} (Montreal, Quebec, Canada)
  \emph{(\bibinfo{series}{ICSE '19})}. \bibinfo{publisher}{IEEE Press},
  \bibinfo{pages}{1039–1049}.
\newblock
\urldef\tempurl%
\url{https://doi.org/10.1109/ICSE.2019.00108}
\showDOI{\tempurl}


\bibitem[\protect\citeauthoryear{Kocaballi, Quiroz, Laranjo, Rezazadegan,
  Kocielnik, Clark, Liao, Park, Moore, and Miner}{Kocaballi
  et~al\mbox{.}}{2020}]%
        {healthcare}
\bibfield{author}{\bibinfo{person}{A.~Baki Kocaballi}, \bibinfo{person}{Juan~C.
  Quiroz}, \bibinfo{person}{Liliana Laranjo}, \bibinfo{person}{Dana
  Rezazadegan}, \bibinfo{person}{Rafal Kocielnik}, \bibinfo{person}{Leigh
  Clark}, \bibinfo{person}{Q.~Vera Liao}, \bibinfo{person}{Sun~Young Park},
  \bibinfo{person}{Robert~J. Moore}, {and} \bibinfo{person}{Adam Miner}.}
  \bibinfo{year}{2020}\natexlab{}.
\newblock \showarticletitle{Conversational Agents for Health and Wellbeing}. In
  \bibinfo{booktitle}{\emph{Extended Abstracts of the 2020 CHI Conference on
  Human Factors in Computing Systems}} (Honolulu, HI, USA)
  \emph{(\bibinfo{series}{CHI EA '20})}. \bibinfo{publisher}{Association for
  Computing Machinery}, \bibinfo{address}{New York, NY, USA},
  \bibinfo{pages}{1–8}.
\newblock
\showISBNx{9781450368193}
\urldef\tempurl%
\url{https://doi.org/10.1145/3334480.3375154}
\showDOI{\tempurl}


\bibitem[\protect\citeauthoryear{Kotz, Balakrishnan, Read, and Vidakovic}{Kotz
  et~al\mbox{.}}{2005}]%
        {kotz2005encyclopedia}
\bibfield{author}{\bibinfo{person}{Samuel Kotz}, \bibinfo{person}{Narayanaswamy
  Balakrishnan}, \bibinfo{person}{Campbell~B Read}, {and}
  \bibinfo{person}{Brani Vidakovic}.} \bibinfo{year}{2005}\natexlab{}.
\newblock \bibinfo{booktitle}{\emph{Encyclopedia of Statistical Sciences,
  Volume 1}}.
\newblock \bibinfo{publisher}{John Wiley \& Sons}.
\newblock


\bibitem[\protect\citeauthoryear{Kriman, Beliaev, Ginsburg, Huang, Kuchaiev,
  Lavrukhin, Leary, Li, and Zhang}{Kriman et~al\mbox{.}}{2020}]%
        {quartznet}
\bibfield{author}{\bibinfo{person}{Samuel Kriman}, \bibinfo{person}{Stanislav
  Beliaev}, \bibinfo{person}{Boris Ginsburg}, \bibinfo{person}{Jocelyn Huang},
  \bibinfo{person}{Oleksii Kuchaiev}, \bibinfo{person}{Vitaly Lavrukhin},
  \bibinfo{person}{Ryan Leary}, \bibinfo{person}{Jason Li}, {and}
  \bibinfo{person}{Yang Zhang}.} \bibinfo{year}{2020}\natexlab{}.
\newblock \showarticletitle{QuartzNet: Deep Automatic Speech Recognition with
  1D Time-Channel Separable Convolutions}. In \bibinfo{booktitle}{\emph{ICASSP
  2022 - 2021 IEEE International Conference on Acoustics, Speech and Signal
  Processing (ICASSP)}}. \bibinfo{pages}{6124–6128}.
\newblock


\bibitem[\protect\citeauthoryear{Këpuska and Bohouta}{Këpuska and
  Bohouta}{2018}]%
        {8301638}
\bibfield{author}{\bibinfo{person}{Veton Këpuska} {and} \bibinfo{person}{Gamal
  Bohouta}.} \bibinfo{year}{2018}\natexlab{}.
\newblock \showarticletitle{Next-generation of virtual personal assistants
  (Microsoft Cortana, Apple Siri, Amazon Alexa and Google Home)}. In
  \bibinfo{booktitle}{\emph{2018 IEEE 8th Annual Computing and Communication
  Workshop and Conference (CCWC)}}. \bibinfo{pages}{99--103}.
\newblock
\urldef\tempurl%
\url{https://doi.org/10.1109/CCWC.2018.8301638}
\showDOI{\tempurl}


\bibitem[\protect\citeauthoryear{Lin and Bilmes}{Lin and Bilmes}{2009a}]%
        {lin2009select}
\bibfield{author}{\bibinfo{person}{Hui Lin} {and} \bibinfo{person}{Jeff
  Bilmes}.} \bibinfo{year}{2009}\natexlab{a}.
\newblock \showarticletitle{How to select a good training-data subset for
  transcription: Submodular active selection for sequences}.
\newblock \bibinfo{journal}{\emph{Proceedings of the Annual Conference of the
  International Speech Communication Association, INTERSPEECH}},
  \bibinfo{pages}{2859--2862}.
\newblock
\urldef\tempurl%
\url{https://doi.org/10.21437/Interspeech.2009-730}
\showDOI{\tempurl}


\bibitem[\protect\citeauthoryear{Lin and Bilmes}{Lin and Bilmes}{2009b}]%
        {LinB09-3}
\bibfield{author}{\bibinfo{person}{Hui Lin} {and} \bibinfo{person}{Jeff
  Bilmes}.} \bibinfo{year}{2009}\natexlab{b}.
\newblock \showarticletitle{How to select a good training-data subset for
  transcription: submodular active selection for sequences}. In
  \bibinfo{booktitle}{\emph{INTERSPEECH 2009, 10th Annual Conference of the
  International Speech Communication Association, Brighton, United Kingdom,
  September 6-10, 2009}}. \bibinfo{publisher}{ISCA},
  \bibinfo{pages}{2859--2862}.
\newblock
\urldef\tempurl%
\url{http://www.isca-speech.org/archive/interspeech_2009/i09_2859.html}
\showURL{%
\tempurl}


\bibitem[\protect\citeauthoryear{Liu, Ott, Goyal, Du, Joshi, Chen, Levy, Lewis,
  Zettlemoyer, and Stoyanov}{Liu et~al\mbox{.}}{2019}]%
        {RoBERTa}
\bibfield{author}{\bibinfo{person}{Yinhan Liu}, \bibinfo{person}{Myle Ott},
  \bibinfo{person}{Naman Goyal}, \bibinfo{person}{Jingfei Du},
  \bibinfo{person}{Mandar Joshi}, \bibinfo{person}{Danqi Chen},
  \bibinfo{person}{Omer Levy}, \bibinfo{person}{Mike Lewis},
  \bibinfo{person}{Luke Zettlemoyer}, {and} \bibinfo{person}{Veselin
  Stoyanov}.} \bibinfo{year}{2019}\natexlab{}.
\newblock \showarticletitle{RoBERTa: {A} Robustly Optimized {BERT} Pretraining
  Approach}.
\newblock \bibinfo{journal}{\emph{CoRR}}  \bibinfo{volume}{abs/1907.11692}
  (\bibinfo{year}{2019}).
\newblock
\showeprint[arXiv]{1907.11692}
\urldef\tempurl%
\url{http://arxiv.org/abs/1907.11692}
\showURL{%
\tempurl}


\bibitem[\protect\citeauthoryear{Ma, Juefei-Xu, Xue, Li, Li, Liu, and Zhao}{Ma
  et~al\mbox{.}}{2019}]%
        {DeepCT}
\bibfield{author}{\bibinfo{person}{Lei Ma}, \bibinfo{person}{Felix Juefei-Xu},
  \bibinfo{person}{Minhui Xue}, \bibinfo{person}{Bo Li}, \bibinfo{person}{Li
  Li}, \bibinfo{person}{Yang Liu}, {and} \bibinfo{person}{Jianjun Zhao}.}
  \bibinfo{year}{2019}\natexlab{}.
\newblock \showarticletitle{DeepCT: Tomographic Combinatorial Testing for Deep
  Learning Systems}. In \bibinfo{booktitle}{\emph{2019 IEEE 26th International
  Conference on Software Analysis, Evolution and Reengineering (SANER)}}.
  \bibinfo{pages}{614--618}.
\newblock
\showISSN{1534-5351}
\urldef\tempurl%
\url{https://doi.org/10.1109/SANER.2019.8668044}
\showDOI{\tempurl}


\bibitem[\protect\citeauthoryear{Ma, Juefei-Xu, Zhang, Sun, Xue, Li, Chen, Su,
  Li, Liu, Zhao, and Wang}{Ma et~al\mbox{.}}{2018}]%
        {DeepGauge}
\bibfield{author}{\bibinfo{person}{Lei Ma}, \bibinfo{person}{Felix Juefei-Xu},
  \bibinfo{person}{Fuyuan Zhang}, \bibinfo{person}{Jiyuan Sun},
  \bibinfo{person}{Minhui Xue}, \bibinfo{person}{Bo Li},
  \bibinfo{person}{Chunyang Chen}, \bibinfo{person}{Ting Su},
  \bibinfo{person}{Li Li}, \bibinfo{person}{Yang Liu}, \bibinfo{person}{Jianjun
  Zhao}, {and} \bibinfo{person}{Yadong Wang}.} \bibinfo{year}{2018}\natexlab{}.
\newblock \showarticletitle{DeepGauge: Multi-Granularity Testing Criteria for
  Deep Learning Systems}. In \bibinfo{booktitle}{\emph{Proceedings of the 33rd
  ACM/IEEE International Conference on Automated Software Engineering}}
  (Montpellier, France) \emph{(\bibinfo{series}{ASE 2018})}.
  \bibinfo{publisher}{Association for Computing Machinery},
  \bibinfo{address}{New York, NY, USA}, \bibinfo{pages}{120–131}.
\newblock
\showISBNx{9781450359375}
\urldef\tempurl%
\url{https://doi.org/10.1145/3238147.3238202}
\showDOI{\tempurl}


\bibitem[\protect\citeauthoryear{Ma, Papadakis, Tsakmalis, Cordy, and Traon}{Ma
  et~al\mbox{.}}{2021}]%
        {ma_tosem}
\bibfield{author}{\bibinfo{person}{Wei Ma}, \bibinfo{person}{Mike Papadakis},
  \bibinfo{person}{Anestis Tsakmalis}, \bibinfo{person}{Maxime Cordy}, {and}
  \bibinfo{person}{Yves~Le Traon}.} \bibinfo{year}{2021}\natexlab{}.
\newblock \showarticletitle{Test Selection for Deep Learning Systems}.
\newblock \bibinfo{journal}{\emph{ACM Transactions on Software Engineering and
  Methodology (TOSEM)}} \bibinfo{volume}{30}, \bibinfo{number}{2}, Article
  \bibinfo{articleno}{13} (\bibinfo{date}{jan} \bibinfo{year}{2021}),
  \bibinfo{numpages}{22}~pages.
\newblock
\showISSN{1049-331X}
\urldef\tempurl%
\url{https://doi.org/10.1145/3417330}
\showDOI{\tempurl}


\bibitem[\protect\citeauthoryear{Malhotra, Bansal, and Ganapathy}{Malhotra
  et~al\mbox{.}}{2019}]%
        {malhotra2019active}
\bibfield{author}{\bibinfo{person}{Karan Malhotra}, \bibinfo{person}{Shubham
  Bansal}, {and} \bibinfo{person}{Sriram Ganapathy}.}
  \bibinfo{year}{2019}\natexlab{}.
\newblock \showarticletitle{Active Learning Methods for Low Resource End-to-End
  Speech Recognition.}. In \bibinfo{booktitle}{\emph{INTERSPEECH}}.
  \bibinfo{pages}{2215--2219}.
\newblock


\bibitem[\protect\citeauthoryear{Mendonça, Candeias, Perdigão, Shulby,
  Toniazzo, Klautau, and Aluísio}{Mendonça et~al\mbox{.}}{2014}]%
        {6947957}
\bibfield{author}{\bibinfo{person}{Gustavo Mendonça}, \bibinfo{person}{Sara
  Candeias}, \bibinfo{person}{Fernando Perdigão}, \bibinfo{person}{Christopher
  Shulby}, \bibinfo{person}{Rean Toniazzo}, \bibinfo{person}{Aldebaro Klautau},
  {and} \bibinfo{person}{Sandra Aluísio}.} \bibinfo{year}{2014}\natexlab{}.
\newblock \showarticletitle{A method for the extraction of phonetically-rich
  triphone sentences}. In \bibinfo{booktitle}{\emph{2014 International
  Telecommunications Symposium (ITS)}}. \bibinfo{pages}{1--5}.
\newblock
\urldef\tempurl%
\url{https://doi.org/10.1109/ITS.2014.6947957}
\showDOI{\tempurl}


\bibitem[\protect\citeauthoryear{Mengge, Yu, Zhang, Liu, Zhang, and
  Wang}{Mengge et~al\mbox{.}}{2020}]%
        {mengge-etal-2020-coarse}
\bibfield{author}{\bibinfo{person}{Xue Mengge}, \bibinfo{person}{Bowen Yu},
  \bibinfo{person}{Zhenyu Zhang}, \bibinfo{person}{Tingwen Liu},
  \bibinfo{person}{Yue Zhang}, {and} \bibinfo{person}{Bin Wang}.}
  \bibinfo{year}{2020}\natexlab{}.
\newblock \showarticletitle{{C}oarse-to-{F}ine {P}re-training for {N}amed
  {E}ntity {R}ecognition}. In \bibinfo{booktitle}{\emph{Proceedings of the 2020
  Conference on Empirical Methods in Natural Language Processing (EMNLP)}}.
  \bibinfo{pages}{6345--6354}.
\newblock


\bibitem[\protect\citeauthoryear{Navarro}{Navarro}{2001}]%
        {10.1145/375360.375365}
\bibfield{author}{\bibinfo{person}{Gonzalo Navarro}.}
  \bibinfo{year}{2001}\natexlab{}.
\newblock \showarticletitle{A Guided Tour to Approximate String Matching}.
\newblock \bibinfo{journal}{\emph{ACM Comput. Surv.}} \bibinfo{volume}{33},
  \bibinfo{number}{1} (\bibinfo{date}{mar} \bibinfo{year}{2001}),
  \bibinfo{pages}{31–88}.
\newblock
\showISSN{0360-0300}
\urldef\tempurl%
\url{https://doi.org/10.1145/375360.375365}
\showDOI{\tempurl}


\bibitem[\protect\citeauthoryear{Ni, Wang, Liu, Leung, Lu, and Ma}{Ni
  et~al\mbox{.}}{2015}]%
        {7178848}
\bibfield{author}{\bibinfo{person}{Chongjia Ni}, \bibinfo{person}{Lei Wang},
  \bibinfo{person}{Haibo Liu}, \bibinfo{person}{Cheung-Chi Leung},
  \bibinfo{person}{Li Lu}, {and} \bibinfo{person}{Bin Ma}.}
  \bibinfo{year}{2015}\natexlab{}.
\newblock \showarticletitle{Submodular data selection with acoustic and
  phonetic features for automatic speech recognition}. In
  \bibinfo{booktitle}{\emph{2015 IEEE International Conference on Acoustics,
  Speech and Signal Processing (ICASSP)}}. \bibinfo{pages}{4629--4633}.
\newblock
\urldef\tempurl%
\url{https://doi.org/10.1109/ICASSP.2015.7178848}
\showDOI{\tempurl}


\bibitem[\protect\citeauthoryear{Pan, Bagherzadeh, Ghaleb, and Briand}{Pan
  et~al\mbox{.}}{2022}]%
        {10.1007/s10664-021-10066-6}
\bibfield{author}{\bibinfo{person}{Rongqi Pan}, \bibinfo{person}{Mojtaba
  Bagherzadeh}, \bibinfo{person}{Taher~A. Ghaleb}, {and}
  \bibinfo{person}{Lionel Briand}.} \bibinfo{year}{2022}\natexlab{}.
\newblock \showarticletitle{Test Case Selection and Prioritization Using
  Machine Learning: A Systematic Literature Review}.
\newblock \bibinfo{journal}{\emph{Empirical Softw. Engg.}}
  \bibinfo{volume}{27}, \bibinfo{number}{2} (\bibinfo{date}{mar}
  \bibinfo{year}{2022}), \bibinfo{numpages}{43}~pages.
\newblock
\showISSN{1382-3256}
\urldef\tempurl%
\url{https://doi.org/10.1007/s10664-021-10066-6}
\showDOI{\tempurl}


\bibitem[\protect\citeauthoryear{Panayotov, Chen, Povey, and
  Khudanpur}{Panayotov et~al\mbox{.}}{2015}]%
        {librispeech}
\bibfield{author}{\bibinfo{person}{Vassil Panayotov}, \bibinfo{person}{Guoguo
  Chen}, \bibinfo{person}{Daniel Povey}, {and} \bibinfo{person}{Sanjeev
  Khudanpur}.} \bibinfo{year}{2015}\natexlab{}.
\newblock \showarticletitle{Librispeech: An ASR corpus based on public domain
  audio books}. In \bibinfo{booktitle}{\emph{2015 IEEE International Conference
  on Acoustics, Speech and Signal Processing (ICASSP)}}.
  \bibinfo{pages}{5206--5210}.
\newblock
\urldef\tempurl%
\url{https://doi.org/10.1109/ICASSP.2015.7178964}
\showDOI{\tempurl}


\bibitem[\protect\citeauthoryear{Pei, Cao, Yang, and Jana}{Pei
  et~al\mbox{.}}{2019}]%
        {DeepXplore}
\bibfield{author}{\bibinfo{person}{Kexin Pei}, \bibinfo{person}{Yinzhi Cao},
  \bibinfo{person}{Junfeng Yang}, {and} \bibinfo{person}{Suman Jana}.}
  \bibinfo{year}{2019}\natexlab{}.
\newblock \showarticletitle{DeepXplore: Automated Whitebox Testing of Deep
  Learning Systems}.
\newblock \bibinfo{journal}{\emph{Commun. ACM}} \bibinfo{volume}{62},
  \bibinfo{number}{11} (\bibinfo{date}{Oct.} \bibinfo{year}{2019}),
  \bibinfo{pages}{137–145}.
\newblock
\showISSN{0001-0782}
\urldef\tempurl%
\url{https://doi.org/10.1145/3361566}
\showDOI{\tempurl}


\bibitem[\protect\citeauthoryear{Rajan, Udeshi, and Chattopadhyay}{Rajan
  et~al\mbox{.}}{2022}]%
        {aequevox}
\bibfield{author}{\bibinfo{person}{Sai~Sathiesh Rajan}, \bibinfo{person}{Sakshi
  Udeshi}, {and} \bibinfo{person}{Sudipta Chattopadhyay}.}
  \bibinfo{year}{2022}\natexlab{}.
\newblock \showarticletitle{AequeVox: Automated Fairness Testing of Speech
  Recognition Systems}. In \bibinfo{booktitle}{\emph{25th International
  Conference on Fundamental Approaches to Software Engineering (FASE)}}.
\newblock


\bibitem[\protect\citeauthoryear{Ramesh, KhudaBukhsh, and Kumar}{Ramesh
  et~al\mbox{.}}{2022}]%
        {bitch}
\bibfield{author}{\bibinfo{person}{Krithika Ramesh},
  \bibinfo{person}{Ashiqur~R. KhudaBukhsh}, {and} \bibinfo{person}{Sumeet
  Kumar}.} \bibinfo{year}{2022}\natexlab{}.
\newblock \bibinfo{title}{'Beach' to 'Bitch': Inadvertent Unsafe Transcription
  of Kids' Content on YouTube}.
\newblock
\newblock


\bibitem[\protect\citeauthoryear{Shi, Yang, He, Xu, and Lo}{Shi
  et~al\mbox{.}}{2022}]%
        {shi2022identifier}
\bibfield{author}{\bibinfo{person}{Jieke Shi}, \bibinfo{person}{Zhou Yang},
  \bibinfo{person}{Junda He}, \bibinfo{person}{Bowen Xu}, {and}
  \bibinfo{person}{David Lo}.} \bibinfo{year}{2022}\natexlab{}.
\newblock \showarticletitle{Can Identifier Splitting Improve Open-Vocabulary
  Language Model of Code?}. In \bibinfo{booktitle}{\emph{2022 IEEE
  International Conference on Software Analysis, Evolution and Reengineering
  (SANER)}}. \bibinfo{publisher}{IEEE Computer Society}.
\newblock


\bibitem[\protect\citeauthoryear{Shinohara}{Shinohara}{2014}]%
        {phonme_rich}
\bibfield{author}{\bibinfo{person}{Yusuke Shinohara}.}
  \bibinfo{year}{2014}\natexlab{}.
\newblock \showarticletitle{A submodular optimization approach to sentence set
  selection}. In \bibinfo{booktitle}{\emph{2014 IEEE International Conference
  on Acoustics, Speech and Signal Processing (ICASSP)}}.
  \bibinfo{pages}{4112--4115}.
\newblock
\urldef\tempurl%
\url{https://doi.org/10.1109/ICASSP.2014.phonme_rich}
\showDOI{\tempurl}


\bibitem[\protect\citeauthoryear{Strayer, Turrill, Coleman, Ortiz, and
  Cooper}{Strayer et~al\mbox{.}}{2014}]%
        {strayer2014measuring}
\bibfield{author}{\bibinfo{person}{David~L Strayer}, \bibinfo{person}{Jonna
  Turrill}, \bibinfo{person}{James~R Coleman}, \bibinfo{person}{Emily~V Ortiz},
  {and} \bibinfo{person}{Joel~M Cooper}.} \bibinfo{year}{2014}\natexlab{}.
\newblock \showarticletitle{Measuring cognitive distraction in the automobile
  II: Assessing in-vehicle voice-based interactive technologies}.
\newblock  (\bibinfo{year}{2014}).
\newblock


\bibitem[\protect\citeauthoryear{Tian, Pei, Jana, and Ray}{Tian
  et~al\mbox{.}}{2018}]%
        {DeepTest}
\bibfield{author}{\bibinfo{person}{Yuchi Tian}, \bibinfo{person}{Kexin Pei},
  \bibinfo{person}{Suman Jana}, {and} \bibinfo{person}{Baishakhi Ray}.}
  \bibinfo{year}{2018}\natexlab{}.
\newblock \showarticletitle{DeepTest: Automated Testing of
  Deep-Neural-Network-Driven Autonomous Cars}. In
  \bibinfo{booktitle}{\emph{Proceedings of the 40th International Conference on
  Software Engineering}} (Gothenburg, Sweden) \emph{(\bibinfo{series}{ICSE
  '18})}. \bibinfo{publisher}{Association for Computing Machinery},
  \bibinfo{address}{New York, NY, USA}, \bibinfo{pages}{303–314}.
\newblock
\showISBNx{9781450356381}
\urldef\tempurl%
\url{https://doi.org/10.1145/3180155.3180220}
\showDOI{\tempurl}


\bibitem[\protect\citeauthoryear{Varadarajan, Yu, Deng, and Acero}{Varadarajan
  et~al\mbox{.}}{2009}]%
        {4960685}
\bibfield{author}{\bibinfo{person}{Balakrishnan Varadarajan},
  \bibinfo{person}{Dong Yu}, \bibinfo{person}{Li Deng}, {and}
  \bibinfo{person}{Alex Acero}.} \bibinfo{year}{2009}\natexlab{}.
\newblock \showarticletitle{Maximizing global entropy reduction for active
  learning in speech recognition}. In \bibinfo{booktitle}{\emph{2009 IEEE
  International Conference on Acoustics, Speech and Signal Processing}}.
  \bibinfo{pages}{4721--4724}.
\newblock
\urldef\tempurl%
\url{https://doi.org/10.1109/ICASSP.2009.4960685}
\showDOI{\tempurl}


\bibitem[\protect\citeauthoryear{Vaswani, Shazeer, Parmar, Uszkoreit, Jones,
  Gomez, Kaiser, and Polosukhin}{Vaswani et~al\mbox{.}}{2017}]%
        {vaswani2017attention}
\bibfield{author}{\bibinfo{person}{Ashish Vaswani}, \bibinfo{person}{Noam
  Shazeer}, \bibinfo{person}{Niki Parmar}, \bibinfo{person}{Jakob Uszkoreit},
  \bibinfo{person}{Llion Jones}, \bibinfo{person}{Aidan~N Gomez},
  \bibinfo{person}{{\L}ukasz Kaiser}, {and} \bibinfo{person}{Illia
  Polosukhin}.} \bibinfo{year}{2017}\natexlab{}.
\newblock \showarticletitle{Attention is all you need}.
\newblock \bibinfo{journal}{\emph{Advances in neural information processing
  systems}}  \bibinfo{volume}{30} (\bibinfo{year}{2017}).
\newblock


\bibitem[\protect\citeauthoryear{Vignesh, Shanmugam, and Murthy}{Vignesh
  et~al\mbox{.}}{2016}]%
        {IndicTTS}
\bibfield{author}{\bibinfo{person}{S.~Rupak Vignesh}, \bibinfo{person}{S.~Aswin
  Shanmugam}, {and} \bibinfo{person}{Hema~A. Murthy}.}
  \bibinfo{year}{2016}\natexlab{}.
\newblock \showarticletitle{Significance of Pseudo-syllables in building better
  acoustic models for Indian English TTS}. In \bibinfo{booktitle}{\emph{2016
  IEEE International Conference on Acoustics, Speech and Signal Processing
  (ICASSP)}}. \bibinfo{pages}{5620--5624}.
\newblock
\urldef\tempurl%
\url{https://doi.org/10.1109/ICASSP.2016.7472753}
\showDOI{\tempurl}


\bibitem[\protect\citeauthoryear{Wei, Liu, Kirchhoff, Bartels, and Bilmes}{Wei
  et~al\mbox{.}}{2014}]%
        {6854213}
\bibfield{author}{\bibinfo{person}{Kai Wei}, \bibinfo{person}{Yuzong Liu},
  \bibinfo{person}{Katrin Kirchhoff}, \bibinfo{person}{Chris Bartels}, {and}
  \bibinfo{person}{Jeff Bilmes}.} \bibinfo{year}{2014}\natexlab{}.
\newblock \showarticletitle{Submodular subset selection for large-scale speech
  training data}. In \bibinfo{booktitle}{\emph{2014 IEEE International
  Conference on Acoustics, Speech and Signal Processing (ICASSP)}}.
  \bibinfo{pages}{3311--3315}.
\newblock
\urldef\tempurl%
\url{https://doi.org/10.1109/ICASSP.2014.6854213}
\showDOI{\tempurl}


\bibitem[\protect\citeauthoryear{Wu and Rajan}{Wu and Rajan}{2021}]%
        {catchme}
\bibfield{author}{\bibinfo{person}{Xiaoliang Wu} {and} \bibinfo{person}{Ajitha
  Rajan}.} \bibinfo{year}{2021}\natexlab{}.
\newblock \bibinfo{title}{Catch Me If You Can: Blackbox Adversarial Attacks on
  Automatic Speech Recognition using Frequency Masking}.
\newblock
\newblock
\urldef\tempurl%
\url{https://doi.org/10.48550/ARXIV.2112.01821}
\showDOI{\tempurl}


\bibitem[\protect\citeauthoryear{Wu, Schuster, Chen, Le, Norouzi, Macherey,
  Krikun, Cao, Gao, Macherey, Klingner, Shah, Johnson, Liu, Kaiser, Gouws,
  Kato, Kudo, Kazawa, Stevens, Kurian, Patil, Wang, Young, Smith, Riesa,
  Rudnick, Vinyals, Corrado, Hughes, and Dean}{Wu et~al\mbox{.}}{2016}]%
        {wordpiece}
\bibfield{author}{\bibinfo{person}{Yonghui Wu}, \bibinfo{person}{Mike
  Schuster}, \bibinfo{person}{Zhifeng Chen}, \bibinfo{person}{Quoc~V. Le},
  \bibinfo{person}{Mohammad Norouzi}, \bibinfo{person}{Wolfgang Macherey},
  \bibinfo{person}{Maxim Krikun}, \bibinfo{person}{Yuan Cao},
  \bibinfo{person}{Qin Gao}, \bibinfo{person}{Klaus Macherey},
  \bibinfo{person}{Jeff Klingner}, \bibinfo{person}{Apurva Shah},
  \bibinfo{person}{Melvin Johnson}, \bibinfo{person}{Xiaobing Liu},
  \bibinfo{person}{Lukasz Kaiser}, \bibinfo{person}{Stephan Gouws},
  \bibinfo{person}{Yoshikiyo Kato}, \bibinfo{person}{Taku Kudo},
  \bibinfo{person}{Hideto Kazawa}, \bibinfo{person}{Keith Stevens},
  \bibinfo{person}{George Kurian}, \bibinfo{person}{Nishant Patil},
  \bibinfo{person}{Wei Wang}, \bibinfo{person}{Cliff Young},
  \bibinfo{person}{Jason Smith}, \bibinfo{person}{Jason Riesa},
  \bibinfo{person}{Alex Rudnick}, \bibinfo{person}{Oriol Vinyals},
  \bibinfo{person}{Greg Corrado}, \bibinfo{person}{Macduff Hughes}, {and}
  \bibinfo{person}{Jeffrey Dean}.} \bibinfo{year}{2016}\natexlab{}.
\newblock \showarticletitle{Google's Neural Machine Translation System:
  Bridging the Gap between Human and Machine Translation}.
\newblock \bibinfo{journal}{\emph{CoRR}}  \bibinfo{volume}{abs/1609.08144}
  (\bibinfo{year}{2016}).
\newblock
\showeprint[arXiv]{1609.08144}
\urldef\tempurl%
\url{http://arxiv.org/abs/1609.08144}
\showURL{%
\tempurl}


\bibitem[\protect\citeauthoryear{Xie, Ma, Juefei-Xu, Xue, Chen, Liu, Zhao, Li,
  Yin, and See}{Xie et~al\mbox{.}}{2019}]%
        {DeepHunter}
\bibfield{author}{\bibinfo{person}{Xiaofei Xie}, \bibinfo{person}{Lei Ma},
  \bibinfo{person}{Felix Juefei-Xu}, \bibinfo{person}{Minhui Xue},
  \bibinfo{person}{Hongxu Chen}, \bibinfo{person}{Yang Liu},
  \bibinfo{person}{Jianjun Zhao}, \bibinfo{person}{Bo Li},
  \bibinfo{person}{Jianxiong Yin}, {and} \bibinfo{person}{Simon See}.}
  \bibinfo{year}{2019}\natexlab{}.
\newblock \showarticletitle{DeepHunter: A Coverage-Guided Fuzz Testing
  Framework for Deep Neural Networks}. In \bibinfo{booktitle}{\emph{Proceedings
  of the 28th ACM SIGSOFT International Symposium on Software Testing and
  Analysis}} (Beijing, China) \emph{(\bibinfo{series}{ISSTA 2019})}.
  \bibinfo{publisher}{Association for Computing Machinery},
  \bibinfo{address}{New York, NY, USA}, \bibinfo{pages}{146–157}.
\newblock
\showISBNx{9781450362245}
\urldef\tempurl%
\url{https://doi.org/10.1145/3293882.3330579}
\showDOI{\tempurl}


\bibitem[\protect\citeauthoryear{Yan, Tao, Liu, Zhai, Ma, Xu, and Zhang}{Yan
  et~al\mbox{.}}{2020}]%
        {FSE_Yan}
\bibfield{author}{\bibinfo{person}{Shenao Yan}, \bibinfo{person}{Guanhong Tao},
  \bibinfo{person}{Xuwei Liu}, \bibinfo{person}{Juan Zhai},
  \bibinfo{person}{Shiqing Ma}, \bibinfo{person}{Lei Xu}, {and}
  \bibinfo{person}{Xiangyu Zhang}.} \bibinfo{year}{2020}\natexlab{}.
\newblock \showarticletitle{Correlations between Deep Neural Network Model
  Coverage Criteria and Model Quality}. In
  \bibinfo{booktitle}{\emph{Proceedings of the 28th ACM Joint Meeting on
  European Software Engineering Conference and Symposium on the Foundations of
  Software Engineering}} (Virtual Event, USA) \emph{(\bibinfo{series}{ESEC/FSE
  2020})}. \bibinfo{publisher}{Association for Computing Machinery},
  \bibinfo{address}{New York, NY, USA}, \bibinfo{pages}{775–787}.
\newblock
\showISBNx{9781450370431}
\urldef\tempurl%
\url{https://doi.org/10.1145/3368089.3409671}
\showDOI{\tempurl}


\bibitem[\protect\citeauthoryear{Yang, Shi, Asyrofi, and Lo}{Yang
  et~al\mbox{.}}{2022a}]%
        {yang2022revisiting}
\bibfield{author}{\bibinfo{person}{Zhou Yang}, \bibinfo{person}{Jieke Shi},
  \bibinfo{person}{Muhammad~Hilmi Asyrofi}, {and} \bibinfo{person}{David Lo}.}
  \bibinfo{year}{2022}\natexlab{a}.
\newblock \showarticletitle{Revisiting Neuron Coverage Metrics and Quality of
  Deep Neural Networks}. In \bibinfo{booktitle}{\emph{2022 IEEE International
  Conference on Software Analysis, Evolution and Reengineering (SANER)}}.
\newblock


\bibitem[\protect\citeauthoryear{Yang, Shi, He, and Lo}{Yang
  et~al\mbox{.}}{2022b}]%
        {10.1145/3510003.3510146}
\bibfield{author}{\bibinfo{person}{Zhou Yang}, \bibinfo{person}{Jieke Shi},
  \bibinfo{person}{Junda He}, {and} \bibinfo{person}{David Lo}.}
  \bibinfo{year}{2022}\natexlab{b}.
\newblock \showarticletitle{Natural Attack for Pre-Trained Models of Code}. In
  \bibinfo{booktitle}{\emph{Proceedings of the 44th International Conference on
  Software Engineering}} (Pittsburgh, Pennsylvania)
  \emph{(\bibinfo{series}{ICSE '22})}. \bibinfo{publisher}{Association for
  Computing Machinery}, \bibinfo{address}{New York, NY, USA},
  \bibinfo{pages}{1482–1493}.
\newblock
\showISBNx{9781450392211}
\urldef\tempurl%
\url{https://doi.org/10.1145/3510003.3510146}
\showDOI{\tempurl}


\bibitem[\protect\citeauthoryear{Yoo and Harman}{Yoo and Harman}{2012}]%
        {testsurvey}
\bibfield{author}{\bibinfo{person}{S. Yoo} {and} \bibinfo{person}{M. Harman}.}
  \bibinfo{year}{2012}\natexlab{}.
\newblock \showarticletitle{Regression Testing Minimization, Selection and
  Prioritization: A Survey}.
\newblock \bibinfo{journal}{\emph{Softw. Test. Verif. Reliab.}}
  \bibinfo{volume}{22}, \bibinfo{number}{2} (\bibinfo{date}{mar}
  \bibinfo{year}{2012}), \bibinfo{pages}{67–120}.
\newblock
\showISSN{0960-0833}
\urldef\tempurl%
\url{https://doi.org/10.1002/stv.430}
\showDOI{\tempurl}


\bibitem[\protect\citeauthoryear{Yoshimura, Yoshida, Matulic, and
  Igarashi}{Yoshimura et~al\mbox{.}}{2019}]%
        {robot_control}
\bibfield{author}{\bibinfo{person}{Naoya Yoshimura}, \bibinfo{person}{Hironori
  Yoshida}, \bibinfo{person}{Fabrice Matulic}, {and} \bibinfo{person}{Takeo
  Igarashi}.} \bibinfo{year}{2019}\natexlab{}.
\newblock \showarticletitle{Extending Discrete Verbal Commands with Continuous
  Speech for Flexible Robot Control}. In \bibinfo{booktitle}{\emph{Extended
  Abstracts of the 2019 CHI Conference on Human Factors in Computing Systems}}
  (Glasgow, Scotland Uk) \emph{(\bibinfo{series}{CHI EA '19})}.
  \bibinfo{publisher}{Association for Computing Machinery},
  \bibinfo{address}{New York, NY, USA}, \bibinfo{pages}{1–6}.
\newblock
\showISBNx{9781450359719}
\urldef\tempurl%
\url{https://doi.org/10.1145/3290607.3312791}
\showDOI{\tempurl}


\bibitem[\protect\citeauthoryear{Zhang, Harman, Ma, and Liu}{Zhang
  et~al\mbox{.}}{2022}]%
        {ml-testing-survey}
\bibfield{author}{\bibinfo{person}{Jie~M. Zhang}, \bibinfo{person}{Mark
  Harman}, \bibinfo{person}{Lei Ma}, {and} \bibinfo{person}{Yang Liu}.}
  \bibinfo{year}{2022}\natexlab{}.
\newblock \showarticletitle{Machine Learning Testing: Survey, Landscapes and
  Horizons}.
\newblock \bibinfo{journal}{\emph{IEEE Transactions on Software Engineering}}
  \bibinfo{volume}{48}, \bibinfo{number}{1} (\bibinfo{year}{2022}),
  \bibinfo{pages}{1--36}.
\newblock
\urldef\tempurl%
\url{https://doi.org/10.1109/TSE.2019.2962027}
\showDOI{\tempurl}


\bibitem[\protect\citeauthoryear{Zhang, Xu, Thung, Haryono, Lo, and
  Jiang}{Zhang et~al\mbox{.}}{2020}]%
        {9240704}
\bibfield{author}{\bibinfo{person}{Ting Zhang}, \bibinfo{person}{Bowen Xu},
  \bibinfo{person}{Ferdian Thung}, \bibinfo{person}{Stefanus~Agus Haryono},
  \bibinfo{person}{David Lo}, {and} \bibinfo{person}{Lingxiao Jiang}.}
  \bibinfo{year}{2020}\natexlab{}.
\newblock \showarticletitle{Sentiment Analysis for Software Engineering: How
  Far Can Pre-trained Transformer Models Go?}. In
  \bibinfo{booktitle}{\emph{2020 IEEE International Conference on Software
  Maintenance and Evolution (ICSME)}}. \bibinfo{pages}{70--80}.
\newblock
\urldef\tempurl%
\url{https://doi.org/10.1109/ICSME46990.2020.00017}
\showDOI{\tempurl}


\bibitem[\protect\citeauthoryear{{Zhao}, {Sonsaat}, {Silpachai}, {Lucic},
  {Chukharev-Hudilainen}, {Levis}, and {Gutierrez-Osuna}}{{Zhao}
  et~al\mbox{.}}{2018}]%
        {l2arctic}
\bibfield{author}{\bibinfo{person}{Guanlong {Zhao}}, \bibinfo{person}{Sinem
  {Sonsaat}}, \bibinfo{person}{Alif {Silpachai}}, \bibinfo{person}{Ivana
  {Lucic}}, \bibinfo{person}{Evgeny {Chukharev-Hudilainen}},
  \bibinfo{person}{John {Levis}}, {and} \bibinfo{person}{Ricardo
  {Gutierrez-Osuna}}.} \bibinfo{year}{2018}\natexlab{}.
\newblock \showarticletitle{L2-ARCTIC: A Non-native English Speech Corpus}. In
  \bibinfo{booktitle}{\emph{Proc. Interspeech}}. \bibinfo{pages}{2783–2787}.
\newblock
\urldef\tempurl%
\url{https://doi.org/10.21437/Interspeech.2018-1110}
\showDOI{\tempurl}


\end{thebibliography}


\begin{thebibliography}{10}
\providecommand{\url}[1]{#1}
\csname url@samestyle\endcsname
\providecommand{\newblock}{\relax}
\providecommand{\bibinfo}[2]{#2}
\providecommand{\BIBentrySTDinterwordspacing}{\spaceskip=0pt\relax}
\providecommand{\BIBentryALTinterwordstretchfactor}{4}
\providecommand{\BIBentryALTinterwordspacing}{\spaceskip=\fontdimen2\font plus
\BIBentryALTinterwordstretchfactor\fontdimen3\font minus
  \fontdimen4\font\relax}
\providecommand{\BIBforeignlanguage}[2]{{%
\expandafter\ifx\csname l@#1\endcsname\relax
\typeout{** WARNING: IEEEtran.bst: No hyphenation pattern has been}%
\typeout{** loaded for the language `#1'. Using the pattern for}%
\typeout{** the default language instead.}%
\else
\language=\csname l@#1\endcsname
\fi
#2}}
\providecommand{\BIBdecl}{\relax}
\BIBdecl

\bibitem{8301638}
V.~Këpuska and G.~Bohouta, ``Next-generation of virtual personal assistants
  (microsoft cortana, apple siri, amazon alexa and google home),'' in
  \emph{2018 IEEE 8th Annual Computing and Communication Workshop and
  Conference (CCWC)}, 2018, pp. 99--103.

\bibitem{robot_control}
\BIBentryALTinterwordspacing
N.~Yoshimura, H.~Yoshida, F.~Matulic, and T.~Igarashi, ``Extending discrete
  verbal commands with continuous speech for flexible robot control,'' in
  \emph{Extended Abstracts of the 2019 CHI Conference on Human Factors in
  Computing Systems}, ser. CHI EA '19.\hskip 1em plus 0.5em minus 0.4em\relax
  New York, NY, USA: Association for Computing Machinery, 2019, p. 1–6.
  [Online]. Available: \url{https://doi.org/10.1145/3290607.3312791}
\BIBentrySTDinterwordspacing

\bibitem{healthcare}
\BIBentryALTinterwordspacing
A.~B. Kocaballi, J.~C. Quiroz, L.~Laranjo, D.~Rezazadegan, R.~Kocielnik,
  L.~Clark, Q.~V. Liao, S.~Y. Park, R.~J. Moore, and A.~Miner, ``Conversational
  agents for health and wellbeing,'' in \emph{Extended Abstracts of the 2020
  CHI Conference on Human Factors in Computing Systems}, ser. CHI EA '20.\hskip
  1em plus 0.5em minus 0.4em\relax New York, NY, USA: Association for Computing
  Machinery, 2020, p. 1–8. [Online]. Available:
  \url{https://doi.org/10.1145/3334480.3375154}
\BIBentrySTDinterwordspacing

\bibitem{strayer2014measuring}
D.~L. Strayer, J.~Turrill, J.~R. Coleman, E.~V. Ortiz, and J.~M. Cooper,
  ``Measuring cognitive distraction in the automobile ii: Assessing in-vehicle
  voice-based interactive technologies,'' 2014.

\bibitem{cooper2014mental}
J.~M. Cooper, H.~Ingebretsen, and D.~L. Strayer, ``Mental workload of common
  voice-based vehicle interactions across six different vehicle systems,''
  2014.

\bibitem{bitch}
K.~Ramesh, A.~R. KhudaBukhsh, and S.~Kumar, ``'beach' to 'bitch': Inadvertent
  unsafe transcription of kids' content on youtube,'' 2022.

\bibitem{crossasr}
M.~H. Asyrofi, F.~Thung, D.~Lo, and L.~Jiang, ``Crossasr: Efficient
  differential testing of automatic speech recognition via text-to-speech,'' in
  \emph{2020 IEEE International Conference on Software Maintenance and
  Evolution (ICSME)}, 2020, pp. 640--650.

\bibitem{crossasrpp}
\BIBentryALTinterwordspacing
M.~H. Asyrofi, Z.~Yang, and D.~Lo, ``Crossasr++: A modular differential testing
  framework for automatic speech recognition,'' in \emph{Proceedings of the
  29th ACM Joint Meeting on European Software Engineering Conference and
  Symposium on the Foundations of Software Engineering}, ser. ESEC/FSE
  2021.\hskip 1em plus 0.5em minus 0.4em\relax New York, NY, USA: Association
  for Computing Machinery, 2021, p. 1575–1579. [Online]. Available:
  \url{https://doi.org/10.1145/3468264.3473124}
\BIBentrySTDinterwordspacing

\bibitem{aequevox}
S.~S. Rajan, S.~Udeshi, and S.~Chattopadhyay, ``Aequevox: Automated fairness
  testing of speech recognition systems,'' in \emph{25th International
  Conference on Fundamental Approaches to Software Engineering (FASE)}, 2022.

\bibitem{ASRTest}
\BIBentryALTinterwordspacing
P.~Ji, Y.~Feng, J.~Liu, Z.~Zhao, and Z.~Chen, ``Asrtest: Automated testing for
  deep-neural-network-driven speech recognition systems,'' in \emph{Proceedings
  of the 31st ACM SIGSOFT International Symposium on Software Testing and
  Analysis}, ser. ISSTA 2022.\hskip 1em plus 0.5em minus 0.4em\relax New York,
  NY, USA: Association for Computing Machinery, 2022, p. 189–201. [Online].
  Available: \url{https://doi.org/10.1145/3533767.3534391}
\BIBentrySTDinterwordspacing

\bibitem{testsurvey}
\BIBentryALTinterwordspacing
S.~Yoo and M.~Harman, ``Regression testing minimization, selection and
  prioritization: A survey,'' \emph{Softw. Test. Verif. Reliab.}, vol.~22,
  no.~2, p. 67–120, mar 2012. [Online]. Available:
  \url{https://doi.org/10.1002/stv.430}
\BIBentrySTDinterwordspacing

\bibitem{icassp2021}
A.~Awasthi, A.~Kansal, S.~Sarawagi, and P.~Jyothi, ``Error-driven fixed-budget
  asr personalization for accented speakers,'' in \emph{ICASSP 2021 - 2021 IEEE
  International Conference on Acoustics, Speech and Signal Processing
  (ICASSP)}, 2021, pp. 7033--7037.

\bibitem{phonme_rich}
Y.~Shinohara, ``A submodular optimization approach to sentence set selection,''
  in \emph{2014 IEEE International Conference on Acoustics, Speech and Signal
  Processing (ICASSP)}, 2014, pp. 4112--4115.

\bibitem{bert}
\BIBentryALTinterwordspacing
J.~Devlin, M.-W. Chang, K.~Lee, and K.~Toutanova, ``{BERT}: Pre-training of
  deep bidirectional transformers for language understanding,'' in
  \emph{Proceedings of the 2019 Conference of the North {A}merican Chapter of
  the Association for Computational Linguistics: Human Language Technologies,
  Volume 1 (Long and Short Papers)}.\hskip 1em plus 0.5em minus 0.4em\relax
  Minneapolis, Minnesota: Association for Computational Linguistics, Jun. 2019,
  pp. 4171--4186. [Online]. Available: \url{https://aclanthology.org/N19-1423}
\BIBentrySTDinterwordspacing

\bibitem{RoBERTa}
\BIBentryALTinterwordspacing
Y.~Liu, M.~Ott, N.~Goyal, J.~Du, M.~Joshi, D.~Chen, O.~Levy, M.~Lewis,
  L.~Zettlemoyer, and V.~Stoyanov, ``Roberta: {A} robustly optimized {BERT}
  pretraining approach,'' \emph{CoRR}, vol. abs/1907.11692, 2019. [Online].
  Available: \url{http://arxiv.org/abs/1907.11692}
\BIBentrySTDinterwordspacing

\bibitem{quartznet}
S.~Kriman, S.~Beliaev, B.~Ginsburg, J.~Huang, O.~Kuchaiev, V.~Lavrukhin,
  R.~Leary, J.~Li, and Y.~Zhang, ``Quartznet: Deep automatic speech recognition
  with 1d time-channel separable convolutions,'' in \emph{ICASSP 2022 - 2021
  IEEE International Conference on Acoustics, Speech and Signal Processing
  (ICASSP)}, 2020, p. 6124–6128.

\bibitem{HuBERT}
W.-N. Hsu, B.~Bolte, Y.-H.~H. Tsai, K.~Lakhotia, R.~Salakhutdinov, and
  A.~Mohamed, ``Hubert: Self-supervised speech representation learning by
  masked prediction of hidden units,'' \emph{IEEE/ACM Transactions on Audio,
  Speech, and Language Processing}, vol.~29, pp. 3451--3460, 2021.

\bibitem{wav2vec2}
A.~Baevski, Y.~Zhou, A.~Mohamed, and M.~Auli, ``wav2vec 2.0: A framework for
  self-supervised learning of speech representations,'' \emph{Advances in
  Neural Information Processing Systems}, vol.~33, pp. 12\,449--12\,460, 2020.

\bibitem{vaswani2017attention}
A.~Vaswani, N.~Shazeer, N.~Parmar, J.~Uszkoreit, L.~Jones, A.~N. Gomez,
  {\L}.~Kaiser, and I.~Polosukhin, ``Attention is all you need,''
  \emph{Advances in neural information processing systems}, vol.~30, 2017.

\bibitem{7178848}
C.~Ni, L.~Wang, H.~Liu, C.-C. Leung, L.~Lu, and B.~Ma, ``Submodular data
  selection with acoustic and phonetic features for automatic speech
  recognition,'' in \emph{2015 IEEE International Conference on Acoustics,
  Speech and Signal Processing (ICASSP)}, 2015, pp. 4629--4633.

\bibitem{DBLP:journals/corr/abs-2203-09829}
\BIBentryALTinterwordspacing
A.~H. Azeemi, I.~A. Qazi, and A.~A. Raza, ``Towards representative subset
  selection for self-supervised speech recognition,'' \emph{CoRR}, vol.
  abs/2203.09829, 2022. [Online]. Available:
  \url{https://doi.org/10.48550/arXiv.2203.09829}
\BIBentrySTDinterwordspacing

\bibitem{919106}
S.~Elbaum, A.~Malishevsky, and G.~Rothermel, ``Incorporating varying test costs
  and fault severities into test case prioritization,'' in \emph{Proceedings of
  the 23rd International Conference on Software Engineering. ICSE 2001}, 2001,
  pp. 329--338.

\bibitem{10.1145/3460319.3464810}
\BIBentryALTinterwordspacing
R.~Cheng, L.~Zhang, D.~Marinov, and T.~Xu, ``Test-case prioritization for
  configuration testing,'' in \emph{Proceedings of the 30th ACM SIGSOFT
  International Symposium on Software Testing and Analysis}, ser. ISSTA
  2021.\hskip 1em plus 0.5em minus 0.4em\relax New York, NY, USA: Association
  for Computing Machinery, 2021, p. 452–465. [Online]. Available:
  \url{https://doi.org/10.1145/3460319.3464810}
\BIBentrySTDinterwordspacing

\bibitem{10.1145/3236024.3236053}
\BIBentryALTinterwordspacing
J.~Chen, Y.~Lou, L.~Zhang, J.~Zhou, X.~Wang, D.~Hao, and L.~Zhang, ``Optimizing
  test prioritization via test distribution analysis,'' in \emph{Proceedings of
  the 2018 26th ACM Joint Meeting on European Software Engineering Conference
  and Symposium on the Foundations of Software Engineering}, ser. ESEC/FSE
  2018.\hskip 1em plus 0.5em minus 0.4em\relax New York, NY, USA: Association
  for Computing Machinery, 2018, p. 656–667. [Online]. Available:
  \url{https://doi.org/10.1145/3236024.3236053}
\BIBentrySTDinterwordspacing

\bibitem{surprise}
\BIBentryALTinterwordspacing
J.~Kim, R.~Feldt, and S.~Yoo, ``Guiding deep learning system testing using
  surprise adequacy,'' in \emph{Proceedings of the 41st International
  Conference on Software Engineering}, ser. ICSE '19.\hskip 1em plus 0.5em
  minus 0.4em\relax IEEE Press, 2019, p. 1039–1049. [Online]. Available:
  \url{https://doi.org/10.1109/ICSE.2019.00108}
\BIBentrySTDinterwordspacing

\bibitem{DeepHunter}
\BIBentryALTinterwordspacing
X.~Xie, L.~Ma, F.~Juefei-Xu, M.~Xue, H.~Chen, Y.~Liu, J.~Zhao, B.~Li, J.~Yin,
  and S.~See, ``Deephunter: A coverage-guided fuzz testing framework for deep
  neural networks,'' in \emph{Proceedings of the 28th ACM SIGSOFT International
  Symposium on Software Testing and Analysis}, ser. ISSTA 2019.\hskip 1em plus
  0.5em minus 0.4em\relax New York, NY, USA: Association for Computing
  Machinery, 2019, p. 146–157. [Online]. Available:
  \url{https://doi.org/10.1145/3293882.3330579}
\BIBentrySTDinterwordspacing

\bibitem{DeepXplore}
\BIBentryALTinterwordspacing
K.~Pei, Y.~Cao, J.~Yang, and S.~Jana, ``Deepxplore: Automated whitebox testing
  of deep learning systems,'' \emph{Commun. ACM}, vol.~62, no.~11, p.
  137–145, Oct. 2019. [Online]. Available:
  \url{https://doi.org/10.1145/3361566}
\BIBentrySTDinterwordspacing

\bibitem{sensei}
\BIBentryALTinterwordspacing
X.~Gao, R.~K. Saha, M.~R. Prasad, and A.~Roychoudhury, ``Fuzz testing based
  data augmentation to improve robustness of deep neural networks,'' in
  \emph{Proceedings of the ACM/IEEE 42nd International Conference on Software
  Engineering}, ser. ICSE '20.\hskip 1em plus 0.5em minus 0.4em\relax New York,
  NY, USA: Association for Computing Machinery, 2020, p. 1147–1158. [Online].
  Available: \url{https://doi.org/10.1145/3377811.3380415}
\BIBentrySTDinterwordspacing

\bibitem{ma_tosem}
\BIBentryALTinterwordspacing
W.~Ma, M.~Papadakis, A.~Tsakmalis, M.~Cordy, and Y.~L. Traon, ``Test selection
  for deep learning systems,'' \emph{ACM Transactions on Software Engineering
  and Methodology (TOSEM)}, vol.~30, no.~2, jan 2021. [Online]. Available:
  \url{https://doi.org/10.1145/3417330}
\BIBentrySTDinterwordspacing

\bibitem{harel2020neuron}
\BIBentryALTinterwordspacing
F.~Harel-Canada, L.~Wang, M.~A. Gulzar, Q.~Gu, and M.~Kim, ``Is neuron coverage
  a meaningful measure for testing deep neural networks?'' in \emph{Proceedings
  of the 28th ACM Joint Meeting on European Software Engineering Conference and
  Symposium on the Foundations of Software Engineering}, ser. ESEC/FSE
  2020.\hskip 1em plus 0.5em minus 0.4em\relax New York, NY, USA: Association
  for Computing Machinery, 2020, p. 851–862. [Online]. Available:
  \url{https://doi.org/10.1145/3368089.3409754}
\BIBentrySTDinterwordspacing

\bibitem{DeepGini}
\BIBentryALTinterwordspacing
Y.~Feng, Q.~Shi, X.~Gao, J.~Wan, C.~Fang, and Z.~Chen, ``Deepgini: Prioritizing
  massive tests to enhance the robustness of deep neural networks,'' in
  \emph{Proceedings of the 29th ACM SIGSOFT International Symposium on Software
  Testing and Analysis}, ser. ISSTA 2020.\hskip 1em plus 0.5em minus
  0.4em\relax New York, NY, USA: Association for Computing Machinery, 2020, p.
  177–188. [Online]. Available: \url{https://doi.org/10.1145/3395363.3397357}
\BIBentrySTDinterwordspacing

\bibitem{hochreiter1997long}
S.~Hochreiter and J.~Schmidhuber, ``Long short-term memory,'' \emph{Neural
  computation}, vol.~9, no.~8, pp. 1735--1780, 1997.

\bibitem{10.1145/375360.375365}
\BIBentryALTinterwordspacing
G.~Navarro, ``A guided tour to approximate string matching,'' \emph{ACM Comput.
  Surv.}, vol.~33, no.~1, p. 31–88, mar 2001. [Online]. Available:
  \url{https://doi.org/10.1145/375360.375365}
\BIBentrySTDinterwordspacing

\bibitem{9240704}
T.~Zhang, B.~Xu, F.~Thung, S.~A. Haryono, D.~Lo, and L.~Jiang, ``Sentiment
  analysis for software engineering: How far can pre-trained transformer models
  go?'' in \emph{2020 IEEE International Conference on Software Maintenance and
  Evolution (ICSME)}, 2020, pp. 70--80.

\bibitem{mengge-etal-2020-coarse}
X.~Mengge, B.~Yu, Z.~Zhang, T.~Liu, Y.~Zhang, and B.~Wang, ``{C}oarse-to-{F}ine
  {P}re-training for {N}amed {E}ntity {R}ecognition,'' in \emph{Proceedings of
  the 2020 Conference on Empirical Methods in Natural Language Processing
  (EMNLP)}, 2020, pp. 6345--6354.

\bibitem{9825884}
\BIBentryALTinterwordspacing
C.~Yang, B.~Xu, J.~Khan, G.~Uddin, D.~Han, Z.~Yang, and D.~Lo, ``Aspect-based
  api review classification: How far can pre-trained transformer model go?'' in
  \emph{2022 IEEE International Conference on Software Analysis, Evolution and
  Reengineering (SANER)}.\hskip 1em plus 0.5em minus 0.4em\relax Los Alamitos,
  CA, USA: IEEE Computer Society, mar 2022, pp. 385--395. [Online]. Available:
  \url{https://doi.ieeecomputersociety.org/10.1109/SANER53432.2022.00054}
\BIBentrySTDinterwordspacing

\bibitem{9796213}
\BIBentryALTinterwordspacing
J.~He, B.~Xu, Z.~Yang, D.~Han, C.~Yang, and D.~Lo, ``Ptm4tag: Sharpening tag
  recommendation of stack overflow posts with pre-trained models,'' in
  \emph{2022 IEEE/ACM 30th International Conference on Program Comprehension
  (ICPC)}.\hskip 1em plus 0.5em minus 0.4em\relax Los Alamitos, CA, USA: IEEE
  Computer Society, may 2022, pp. 1--11. [Online]. Available:
  \url{https://doi.ieeecomputersociety.org/10.1145/3524610.3527897}
\BIBentrySTDinterwordspacing

\bibitem{10.1145/3551349.3560421}
\BIBentryALTinterwordspacing
C.~Yang, B.~Xu, F.~Thung, Y.~Shi, T.~Zhang, Z.~Yang, X.~Zhou, J.~Shi, J.~He,
  D.~Han, and D.~Lo, ``Answer summarization for technical queries: Benchmark
  and new approach,'' in \emph{Proceedings of the 37th IEEE/ACM International
  Conference on Automated Software Engineering}, ser. ASE '22.\hskip 1em plus
  0.5em minus 0.4em\relax New York, NY, USA: Association for Computing
  Machinery, 2023. [Online]. Available:
  \url{https://doi.org/10.1145/3551349.3560421}
\BIBentrySTDinterwordspacing

\bibitem{10.1145/3551349.3556964}
\BIBentryALTinterwordspacing
J.~Shi, Z.~Yang, B.~Xu, H.~J. Kang, and D.~Lo, ``Compressing pre-trained models
  of code into 3 mb,'' in \emph{Proceedings of the 37th IEEE/ACM International
  Conference on Automated Software Engineering}, ser. ASE '22.\hskip 1em plus
  0.5em minus 0.4em\relax New York, NY, USA: Association for Computing
  Machinery, 2023. [Online]. Available:
  \url{https://doi.org/10.1145/3551349.3556964}
\BIBentrySTDinterwordspacing

\bibitem{shi2022identifier}
J.~Shi, Z.~Yang, J.~He, B.~Xu, and D.~Lo, ``Can identifier splitting improve
  open-vocabulary language model of code?'' in \emph{2022 IEEE International
  Conference on Software Analysis, Evolution and Reengineering (SANER)}.\hskip
  1em plus 0.5em minus 0.4em\relax IEEE Computer Society, 2022.

\bibitem{wordpiece}
\BIBentryALTinterwordspacing
Y.~Wu, M.~Schuster, Z.~Chen, Q.~V. Le, M.~Norouzi, W.~Macherey, M.~Krikun,
  Y.~Cao, Q.~Gao, K.~Macherey, J.~Klingner, A.~Shah, M.~Johnson, X.~Liu,
  L.~Kaiser, S.~Gouws, Y.~Kato, T.~Kudo, H.~Kazawa, K.~Stevens, G.~Kurian,
  N.~Patil, W.~Wang, C.~Young, J.~Smith, J.~Riesa, A.~Rudnick, O.~Vinyals,
  G.~Corrado, M.~Hughes, and J.~Dean, ``Google's neural machine translation
  system: Bridging the gap between human and machine translation,''
  \emph{CoRR}, vol. abs/1609.08144, 2016. [Online]. Available:
  \url{http://arxiv.org/abs/1609.08144}
\BIBentrySTDinterwordspacing

\bibitem{librispeech}
V.~Panayotov, G.~Chen, D.~Povey, and S.~Khudanpur, ``Librispeech: An asr corpus
  based on public domain audio books,'' in \emph{2015 IEEE International
  Conference on Acoustics, Speech and Signal Processing (ICASSP)}, 2015, pp.
  5206--5210.

\bibitem{common-voice}
\BIBentryALTinterwordspacing
R.~Ardila, M.~Branson, K.~Davis, M.~Kohler, J.~Meyer, M.~Henretty, R.~Morais,
  L.~Saunders, F.~Tyers, and G.~Weber, ``\BIBforeignlanguage{English}{Common
  voice: A massively-multilingual speech corpus},'' in
  \emph{\BIBforeignlanguage{English}{Proceedings of the 12th Language Resources
  and Evaluation Conference}}.\hskip 1em plus 0.5em minus 0.4em\relax
  Marseille, France: European Language Resources Association, May 2020, pp.
  4218--4222. [Online]. Available:
  \url{https://aclanthology.org/2020.lrec-1.520}
\BIBentrySTDinterwordspacing

\bibitem{Libri-Light}
J.~Kahn, M.~Rivière, W.~Zheng, E.~Kharitonov, Q.~Xu, P.~Mazaré, J.~Karadayi,
  V.~Liptchinsky, R.~Collobert, C.~Fuegen, T.~Likhomanenko, G.~Synnaeve,
  A.~Joulin, A.~Mohamed, and E.~Dupoux, ``Libri-light: A benchmark for asr with
  limited or no supervision,'' in \emph{ICASSP 2020 - 2020 IEEE International
  Conference on Acoustics, Speech and Signal Processing (ICASSP)}, 2020, pp.
  7669--7673.

\bibitem{IndicTTS}
S.~R. Vignesh, S.~A. Shanmugam, and H.~A. Murthy, ``Significance of
  pseudo-syllables in building better acoustic models for indian english tts,''
  in \emph{2016 IEEE International Conference on Acoustics, Speech and Signal
  Processing (ICASSP)}, 2016, pp. 5620--5624.

\bibitem{l2arctic}
\BIBentryALTinterwordspacing
G.~{Zhao}, S.~{Sonsaat}, A.~{Silpachai}, I.~{Lucic}, E.~{Chukharev-Hudilainen},
  J.~{Levis}, and R.~{Gutierrez-Osuna}, ``L2-arctic: A non-native english
  speech corpus,'' in \emph{Proc. Interspeech}, 2018, p. 2783–2787. [Online].
  Available: \url{http://dx.doi.org/10.21437/Interspeech.2018-1110}
\BIBentrySTDinterwordspacing

\bibitem{10.1007/s10664-021-10066-6}
\BIBentryALTinterwordspacing
R.~Pan, M.~Bagherzadeh, T.~A. Ghaleb, and L.~Briand, ``Test case selection and
  prioritization using machine learning: A systematic literature review,''
  \emph{Empirical Softw. Engg.}, vol.~27, no.~2, mar 2022. [Online]. Available:
  \url{https://doi.org/10.1007/s10664-021-10066-6}
\BIBentrySTDinterwordspacing

\bibitem{6947957}
G.~Mendonça, S.~Candeias, F.~Perdigão, C.~Shulby, R.~Toniazzo, A.~Klautau,
  and S.~Aluísio, ``A method for the extraction of phonetically-rich triphone
  sentences,'' in \emph{2014 International Telecommunications Symposium (ITS)},
  2014, pp. 1--5.

\bibitem{kotz2005encyclopedia}
S.~Kotz, N.~Balakrishnan, C.~B. Read, and B.~Vidakovic, \emph{Encyclopedia of
  Statistical Sciences, Volume 1}.\hskip 1em plus 0.5em minus 0.4em\relax John
  Wiley \& Sons, 2005.

\bibitem{guilford1950fundamental}
J.~P. Guilford, ``Fundamental statistics in psychology and education,'' 1950.

\bibitem{lin2009select}
H.~Lin and J.~Bilmes, ``How to select a good training-data subset for
  transcription: Submodular active selection for sequences,'' 09 2009, pp.
  2859--2862.

\bibitem{6854213}
K.~Wei, Y.~Liu, K.~Kirchhoff, C.~Bartels, and J.~Bilmes, ``Submodular subset
  selection for large-scale speech training data,'' in \emph{2014 IEEE
  International Conference on Acoustics, Speech and Signal Processing
  (ICASSP)}, 2014, pp. 3311--3315.

\bibitem{ml-testing-survey}
J.~M. Zhang, M.~Harman, L.~Ma, and Y.~Liu, ``Machine learning testing: Survey,
  landscapes and horizons,'' \emph{IEEE Transactions on Software Engineering},
  vol.~48, no.~1, pp. 1--36, 2022.

\bibitem{DeepCruiser}
\BIBentryALTinterwordspacing
X.~Du, X.~Xie, Y.~Li, L.~Ma, J.~Zhao, and Y.~Liu, ``Deepcruiser: Automated
  guided testing for stateful deep learning systems,'' \emph{CoRR}, vol.
  abs/1812.05339, 2018. [Online]. Available:
  \url{http://arxiv.org/abs/1812.05339}
\BIBentrySTDinterwordspacing

\bibitem{asrevolve}
M.~H. Asyrofi, Z.~Yang, J.~Shi, C.~W. Quan, and D.~Lo, ``Can differential
  testing improve automatic speech recognition systems?'' in \emph{2021 IEEE
  International Conference on Software Maintenance and Evolution (ICSME)},
  2021, pp. 674--678.

\bibitem{catchme}
\BIBentryALTinterwordspacing
X.~Wu and A.~Rajan, ``Catch me if you can: Blackbox adversarial attacks on
  automatic speech recognition using frequency masking,'' 2021. [Online].
  Available: \url{https://arxiv.org/abs/2112.01821}
\BIBentrySTDinterwordspacing

\bibitem{4960685}
B.~Varadarajan, D.~Yu, L.~Deng, and A.~Acero, ``Maximizing global entropy
  reduction for active learning in speech recognition,'' in \emph{2009 IEEE
  International Conference on Acoustics, Speech and Signal Processing}, 2009,
  pp. 4721--4724.

\bibitem{confidence}
\BIBentryALTinterwordspacing
D.~Hakkani-T\"{u}r, G.~Riccardi, and G.~Tur, ``An active approach to spoken
  language processing,'' \emph{ACM Trans. Speech Lang. Process.}, vol.~3,
  no.~3, p. 1–31, oct 2006. [Online]. Available:
  \url{https://doi.org/10.1145/1177055.1177056}
\BIBentrySTDinterwordspacing

\bibitem{malhotra2019active}
K.~Malhotra, S.~Bansal, and S.~Ganapathy, ``Active learning methods for low
  resource end-to-end speech recognition.'' in \emph{INTERSPEECH}, 2019, pp.
  2215--2219.

\bibitem{LinB09-3}
\BIBentryALTinterwordspacing
H.~Lin and J.~Bilmes, ``How to select a good training-data subset for
  transcription: submodular active selection for sequences,'' in
  \emph{INTERSPEECH 2009, 10th Annual Conference of the International Speech
  Communication Association, Brighton, United Kingdom, September 6-10,
  2009}.\hskip 1em plus 0.5em minus 0.4em\relax ISCA, 2009, pp. 2859--2862.
  [Online]. Available:
  \url{http://www.isca-speech.org/archive/interspeech_2009/i09_2859.html}
\BIBentrySTDinterwordspacing

\bibitem{biasfinder}
M.~H. Asyrofi, Z.~Yang, I.~N.~B. Yusuf, H.~J. Kang, F.~Thung, and D.~Lo,
  ``Biasfinder: Metamorphic test generation to uncover bias for sentiment
  analysis systems,'' \emph{IEEE Transactions on Software Engineering}, pp.
  1--1, 2021.

\bibitem{10.1145/3510003.3510146}
\BIBentryALTinterwordspacing
Z.~Yang, J.~Shi, J.~He, and D.~Lo, ``Natural attack for pre-trained models of
  code,'' in \emph{Proceedings of the 44th International Conference on Software
  Engineering}, ser. ICSE '22.\hskip 1em plus 0.5em minus 0.4em\relax New York,
  NY, USA: Association for Computing Machinery, 2022, p. 1482–1493. [Online].
  Available: \url{https://doi.org/10.1145/3510003.3510146}
\BIBentrySTDinterwordspacing

\bibitem{10.1145/3564625.3564636}
\BIBentryALTinterwordspacing
C.~Gong, Z.~Yang, Y.~Bai, J.~Shi, A.~Sinha, B.~Xu, D.~Lo, X.~Hou, and G.~Fan,
  ``Curiosity-driven and victim-aware adversarial policies,'' in
  \emph{Proceedings of the 38th Annual Computer Security Applications
  Conference}, ser. ACSAC '22.\hskip 1em plus 0.5em minus 0.4em\relax New York,
  NY, USA: Association for Computing Machinery, 2022, p. 186–200. [Online].
  Available: \url{https://doi.org/10.1145/3564625.3564636}
\BIBentrySTDinterwordspacing

\bibitem{DeepTest}
\BIBentryALTinterwordspacing
Y.~Tian, K.~Pei, S.~Jana, and B.~Ray, ``Deeptest: Automated testing of
  deep-neural-network-driven autonomous cars,'' in \emph{Proceedings of the
  40th International Conference on Software Engineering}, ser. ICSE '18.\hskip
  1em plus 0.5em minus 0.4em\relax New York, NY, USA: Association for Computing
  Machinery, 2018, p. 303–314. [Online]. Available:
  \url{https://doi.org/10.1145/3180155.3180220}
\BIBentrySTDinterwordspacing

\bibitem{DeepGauge}
\BIBentryALTinterwordspacing
L.~Ma, F.~Juefei-Xu, F.~Zhang, J.~Sun, M.~Xue, B.~Li, C.~Chen, T.~Su, L.~Li,
  Y.~Liu, J.~Zhao, and Y.~Wang, ``Deepgauge: Multi-granularity testing criteria
  for deep learning systems,'' in \emph{Proceedings of the 33rd ACM/IEEE
  International Conference on Automated Software Engineering}, ser. ASE
  2018.\hskip 1em plus 0.5em minus 0.4em\relax New York, NY, USA: Association
  for Computing Machinery, 2018, p. 120–131. [Online]. Available:
  \url{https://doi.org/10.1145/3238147.3238202}
\BIBentrySTDinterwordspacing

\bibitem{DeepCT}
\BIBentryALTinterwordspacing
L.~Ma, F.~Juefei-Xu, M.~Xue, B.~Li, L.~Li, Y.~Liu, and J.~Zhao, ``Deepct:
  Tomographic combinatorial testing for deep learning systems,'' in \emph{2019
  IEEE 26th International Conference on Software Analysis, Evolution and
  Reengineering (SANER)}, Feb 2019, pp. 614--618. [Online]. Available:
  \url{https://doi.org/10.1109/SANER.2019.8668044}
\BIBentrySTDinterwordspacing

\bibitem{yang2022revisiting}
Z.~Yang, J.~Shi, M.~H. Asyrofi, and D.~Lo, ``Revisiting neuron coverage metrics
  and quality of deep neural networks,'' in \emph{2022 IEEE International
  Conference on Software Analysis, Evolution and Reengineering (SANER)}, 2022.

\bibitem{FSE_Yan}
\BIBentryALTinterwordspacing
S.~Yan, G.~Tao, X.~Liu, J.~Zhai, S.~Ma, L.~Xu, and X.~Zhang, ``Correlations
  between deep neural network model coverage criteria and model quality,'' in
  \emph{Proceedings of the 28th ACM Joint Meeting on European Software
  Engineering Conference and Symposium on the Foundations of Software
  Engineering}, ser. ESEC/FSE 2020.\hskip 1em plus 0.5em minus 0.4em\relax New
  York, NY, USA: Association for Computing Machinery, 2020, p. 775–787.
  [Online]. Available: \url{https://doi.org/10.1145/3368089.3409671}
\BIBentrySTDinterwordspacing

\bibitem{ICECCS}
Y.~Dong, P.~Zhang, J.~Wang, S.~Liu, J.~Sun, J.~Hao, X.~Wang, L.~Wang, J.~Dong,
  and T.~Dai, ``An empirical study on correlation between coverage and
  robustness for deep neural networks,'' in \emph{2020 25th International
  Conference on Engineering of Complex Computer Systems (ICECCS)}, 2020, pp.
  73--82.

\bibitem{hu_tosem}
Q.~Hu, Y.~Guo, M.~Cordy, X.~Xie, L.~Ma, M.~Papadakis, and Y.~Le~Traon, ``An
  empirical study on data distribution-aware test selection for deep learning
  enhancement,'' \emph{ACM Transactions on Software Engineering and Methodology
  (TOSEM)}, 2022.

\end{thebibliography}

\IEEEdisplaynontitleabstractindextext

\end{document}